\documentclass[epjST]{svjour}

\usepackage{natbib}
\usepackage{amsmath}
\usepackage{overpic}
\usepackage{physics}
\usepackage{siunitx}
\usepackage{graphicx}
\usepackage{color}
\usepackage{tikz}

\newcommand{\be}{\begin{equation}}
\newcommand{\ee}{\end{equation}}
\newcommand{\ba}{\begin{eqnarray}}
\newcommand{\ea}{\end{eqnarray}}

\newcommand{\rRe}{\mathrm{Re}}
\newcommand{\U}{\abs{\vb{U}}}
\newcommand{\vel}{\vb{u}}
\newcommand{\fst}{\vb{f}_{\sigma}}
\newcommand{\mi}{\vec{\textbf{n}}}

\newcommand{\Ca}{\mathrm{Ca}}

\newcommand\etal{{\em et al.}~}

\newcommand{\redline}{\raisebox{2pt}{\tikz{\draw[-,red,line width = 1pt](0,0) -- (3mm,0);}}}

\newcommand{\blueline}{\raisebox{2pt}{\tikz{\draw[-,blue,line width = 1pt](0,0) -- (3mm,0);}}}

\newcommand{\greenline}{\raisebox{2pt}{\tikz{\draw[-,green,line width = 1pt](0,0) -- (3mm,0);}}}

\begin{document}
\title{Pore-scale direct numerical simulation of Haines jumps in a porous media model}
\author{Adam O'Brien\inst{1}\and Shahriar Afkhami\inst{2} \fnmsep\thanks{\email{shahriar.afkhami@njit.edu}} \and 
	         Markus Bussmann\inst{1}}
\institute{Department of Mechanical and Industrial Engineering, University of Toronto, Toronto, ON, Canada M5S 3G8 \and 
               Department of Mathematical Sciences, New Jersey Institute of Technology, Newark, NJ, USA 07102}
\abstract{
Direct numerical simulations are presented for a porous media model consisting  of two immiscible fluids, an invading and defending phase, in a two-dimensional micro-geometry filled with randomly sized and randomly distributed cylinders. First, interface instability and penetration modes are studied when varying the wetting features of a single pore in the porous medium. It is found that the displacement patterns not only change with the capillary number,
as previously observed, but also are a function of the contact angle, even for a viscosity ratio of unity. This is an important conclusion suggesting that capillary number and viscosity ratio alone cannot completely describe the pore-scale displacement. Second, rapid pore-scale displacement is considered, where the displacements are accompanied by sudden interface jumps
from one site to another, known as Haines jumps. The characteristic time and length scales of a Haines jump are examined to better understand the transient dynamics of the jump. We then focus on analyzing the Haines jump in a simple pore configuration where cylinders of equal size are placed at the vertices of equilateral triangles. We use this geometry to provide more insight into the effect of the contact angle at which the Haines jump is predicted. 
} 
\maketitle
\section{Introduction}
\label{sec:intro}

The flow of immiscible fluids in porous media has applications in subsurface water flows, tar sands oil production, and $\text{CO}_2$ sequestration, to name a few \citep{Perazzo2018}.  At the micro-scale, where the capillary forces are dominant, the wetting effects can play a significant role in the displacement dynamics. For example, when a fluid is pushed into
a porous medium, displacing an immiscible fluid, the distribution of fluids depends on wetting properties of the medium.
In the pioneering work of Lenormand, the displacement patterns of a wetting fluid, which is known as the imbibition scenario, 
and a non-wetting fluid, known as the drainage scenario, are studied, showing that    
in slow displacement drainage, fingering like patterns emerge \citep{Lenormand1990}. This 
condition can lead to the invading phase not being able to completely drain the other (defending) phase, 
resulting in some of the displaced phase becoming trapped in the porous medium
\citep{Blunt95,Blunt2016,Singh2019}. The  trapping can be important as it affects the saturation
of the invading liquid which limits the extent to which the defending fluid can be displaced from the medium \citep{Weitz18,Zacharoudiou2018}.



The study of the influence of the contact angle on different displacement patterns is important, as it is also one of the key factors in defining the flow behavior, and therefore the resulting fluid distribution within a porous media \citep{Zhao2016,Singh2019,juanes2019}.
In the pioneering work of Cieplak and Robbins \citep{Cieplak1988,Cieplak1990},  a two-dimensional quasistatic numerical model was developed to describe wetting conditions.
However, the approach in \citep{Cieplak1988,Cieplak1990} simplifies the interface representation as circular arcs connecting the solid geometries,
which in their work are two-dimensional array of disks with random radii. Recent particle-based simulations account for wall wettability in two-dimensional
\citep{Holtzman2015} and quasi-two-dimensional \citep{Jung2016} models. However, particle-based simulations present difficulties in enforcing 
the incompressibility in very viscous flows. Nonetheless, all the studies above consistently show the transition from stable front displacement 
at low contact angles to capillary fingering at high contact angles. 

Pore-scale models have increasingly become a viable tool for predicting 
two-phase flows through porous media, mainly due to recent improved developments;
see \citep{Zhao2019} for a comparison of a variety of pore-scale models.
Direct numerical simulation techniques based on solving the full Navier-Stokes equations provide a robust and high accuracy solution
for pore-scale analysis of multiphase  flows  in  porous  media.   
However, a great challenge in pore-scale direct numerical simulations is the inclusion of 
wetting effects into the  numerical model through the contact angle boundary condition.
There exist direct numerical studies on pore-scale modeling of
wettability effects, such as the phase-field simulation in \citep{BASIRAT2017} and the lattice Boltzmann simulation in \citep{LIU2014}. 
Here we present the results of a direct numerical method that is based on an  interface capturing approach. 
While the majority of previous (Eulerian) interface capturing methods include wetting on body-fitted meshes 
(see e.g. \citep{Bussmann99,AFKHAMI2009a} and for a review \citep{Sui2014}), less progress has been 
made to develop numerical methods that handle arbitrary geometries. To that end, one attractive approach is 
the Immersed Boundary Method (IBM) \citep{Peskin1972,Mittal2005}, because of its ability to model complex geometries, 
and its potential for modeling systems of solid bodies with arbitrary relative motion
\citep{OBrien2019}. Here we use an accurate and robust numerical model developed in 
\citep{OBrien2018} and later extended in \cite{OBrien2019} for combining wetting dynamics, in a 
Volume-of-Fluid (VOF) framework, and the IBM.

The result we present here is first of its kind in that it uses 
a combined VOF/IBM pore-scale direct numerical simulations to quantify the effects of the capillarity, characterized by the capillary number, $\Ca$, and wetting, characterized by the contact angle, $\theta$,
on the displacement phenomena in a porous media model. 
We examine how the wetting controls displacement patterns, and show the crossover from a stable propagating front to a fractal pattern, namely a fingering instability. Our work extends on the literature by identifying the displacement regimes as functions of independently varying $\Ca$ and $\theta$. We illustrate the interface patterns as we change the contact angle, and analyze the instabilities in terms of the growth mechanisms and the rapid movement of interfaces between pores. Based on our results, we describe three main patterns and the crossover from fingering to stable flow as a function of $\Ca$ and $\theta$, at a viscosity ratio of $1$. We show that decreasing the contact angle results in a change of the flow behavior from strong fingering to stabilized front propagation, and that the contact angle effect diminishes as $\Ca$ increases.
The threshold for this  transition is related to the contact angle amongst other parameters.
For wetting invading fluids, the front propagates rather smoothly, draining all of the displaced phase, while a non-wetting invading fluid can lead to residual fluid. Our simulations reveal how post-fingering production flux is 
different above a critical contact angle from the one below the critical contact angle where the front 
is stable, for sufficiently small $\Ca$, and that the production flux is similar when $\Ca$ is sufficiently large, 
irrespective of the contact angle. Our study
therefore allows the quantification of the flow mechanisms when including the wetting effects.
We note that we do not discuss here the effects of gravity, as the capillary forces dominate the gravitational forces,
and so the buoyancy effect is secondary (see e.g.~\cite{Hilfer96} for more discussion). We also note
that an important characteristic of the liquid--solid interaction is the contact angle hysteresis, 
caused mainly due to surface roughness and heterogeneity. While to avoid further complications, we ignore
the effects, it is shown in \cite{Helland2007} that if contact angle hysteresis is assumed,
major displacement mechanisms still remain unaltered. 

We also pay particular attention to sharp interfacial jumps during the displacement, known as Haines jumps \cite{haines1930,Cieplak1988,Lenormand1990}.
This phenomena  can lead to a highly discontinuous rate at which the front progresses, identified as bursting, when it would otherwise progress slowly and continuously. Here we numerically investigate a two-dimensional invading fluid displacing another fluid in an arrangement of circular disks.
We show the influence of the Haines jump on the propagation of the front, which is accompanied by a sharp pressure oscillation. We further investigate the effects of the interfacial tension 
and the wettability on the jump transient dynamics. Finally, we study the frequency of Haines jumps as a function of $\theta$
and $\Ca$. This information can be useful to predict the residual saturation of the displaced phase, which is a measure of the
displacement efficiency.

\section {Computational Setup}  
\label{sec:computational_setup}

In this work, the flow equations are solved numerically and the pore geometry is modeled using an Immersed Boundary (IB) method \cite{OBrien2018} in place of body-fitted meshes. The incompressible Navier-Stokes equations for a two-component fluid system are solved, where the interface is tracked using a Volume-of-Fluid (VOF) method. A contact line boundary condition is imposed on surfaces of the immersed boundaries. This approach has many advantages; in particular it greatly simplifies the inclusion of complex domain geometries, as the need to produce a body-fitted mesh is eliminated.

The Navier-Stokes equations for a two component fluid include the momentum equation,

\begin{equation}
	\pdv{\rho\vel}{t} + \div{\rho\vel\vel} = - \grad p + \div\mu\qty(\grad\vel + \grad\vel^\text{T})  + \fst
	\label{eq:momentum}
\end{equation}

\noindent
and the mass conservation equation,

\begin{equation}
	\pdv{\rho}{t} + \div\rho\vel = 0
	\label{eq:continuity}
\end{equation}

\noindent
where $\rho$ is the fluid density, $\mu$ is the fluid viscosity, $\vel$ is the fluid velocity, $p$ is the pressure, and $\fst$ is a surface tension body force. 


For an incompressible flow consisting of two immiscible phases, the density is expressed as
$\rho(\gamma) = \rho_1 + \gamma\qty(\rho_2-\rho_1)$, while the viscosity is computed from harmonic averaging,
$\frac{\rho(\gamma)}{\mu(\gamma)} = \frac{\rho_1}{\mu_1} + \gamma\qty(\frac{\rho_2}{\mu_2}-\frac{\rho_1}{\mu_1})$,
where $\gamma$ is the fluid volume fraction governed by the VOF advection equation,

\begin{equation}
	\pdv{\gamma}{t} + \div\vel\gamma = 0
	\label{eq:vof}
\end{equation}

\noindent
which is solved using a suitable numerical scheme that limits numerical diffusion in order to maintain a sharp interface between the two fluids. In this work, the Compressive Interface Capturing Scheme for Arbitrary Meshes (CICSAM) scheme of Ubbink and Issa \cite{Ubbink1999} is used to solve the discretized form of Eq.\ \eqref{eq:vof}. The computation of $\fst$ is carried out using the Continuum Surface Force (CSF) method of Brackbill \etal \cite{Brackbill1992}, which models surface tension as

\begin{equation}
	\fst = \sigma \kappa \grad\gamma
\end{equation}

\noindent
where $\sigma$ is the constant surface tension coefficient, and $\kappa$ is the interface curvature,
computed as

\begin{equation}
	\kappa = \div{\mi}
	\label{eq:div_m}
\end{equation}
\noindent
where $\mi$ is the interface unit normal vector, defined as

\begin{equation}
	\mi = -\frac{\grad \tilde{\gamma}}{\left\|\grad \tilde{\gamma}\right\|}
	\label{eq:interface_normal}
\end{equation}
\noindent
where $\tilde{\gamma}$ is the smoothed volume fraction field.
The contact angle is imposed through the computation of the interface curvature near the IB;
see \cite{OBrien2019} for details. Briefly, in order to compute the interface curvature in cells neighboring an IB cell, 
the orientation of the interface normal at the contact line, determined from the prescribed contact angle,
is used for the calculation in Eq.\ \eqref{eq:div_m}, instead of a value of $\mi$ calculated from Eq.\ \eqref{eq:interface_normal}.
To compute $\grad\gamma$ at each point where the numerical stencil intersects the IB, 
ghost-cell values of $\gamma$ are evaluated using a bilinear interpolation and a least-squares stencil to reflect the contact
angle boundary condition.

Finally, to close the system of equations, the divergence-free constraint is imposed on the velocity field, which has the form

\begin{equation}
	\div\vel = 0
	\label{eq:div_free}
\end{equation}

Eqs.\ \eqref{eq:momentum} and \eqref{eq:div_free} are solved together using a fractional-step projection method \cite{chorin1967}, based on the balanced-force algorithm of Francois \etal \cite{Francois2006}. The details of the discretization are omitted here, and can be found in \cite{OBrien2018}. For the simulations reported in this work, Eq.\ \eqref{eq:momentum} is non-dimensionalized as follows,

\begin{equation}
	\Re\qty(\pdv{\lambda_\rho\va{u^*}}{t^*} + \grad^*\cdot{\lambda_\rho\vel^*\vel^*}) = - \grad^* p^* + \grad^*\cdot\lambda_\mu\qty(\grad^*\vel^* + {\grad^*\vel^*}^\text{T})  + \frac{1}{\Ca}\kappa^*\grad^*\gamma
	\label{eq:non-dimensional_momentum}
\end{equation}

\noindent
where the superscript $^*$ denotes a non-dimensional quantity, $\lambda_\rho$ is the density ratio,  $\lambda_\mu$ is the viscosity ratio, and the Reynolds number $\Re$ and Capillary number $\Ca$ are defined as follows,

\begin{equation*}
	\Re = \frac{\rho\U\overline{D}}{\mu}
\end{equation*}
\begin{equation*}
	\Ca = \frac{\mu\U}{\sigma}
\end{equation*}

\noindent
where $\U$ is the characteristic velocity, taken here to be the bulk flow velocity. A no-slip boundary condition is employed on the IB surfaces. It is noted that when using a no-slip boundary condition, a stress singularity is present at the contact line which can lead to results being dependent on the grid size. This phenomenon is discussed in detail by Afkhami \etal in \cite{AFKHAMI2009a,Afkhami2018}.
Although the dynamic contact line models developed in \cite{AFKHAMI2009a,Afkhami2018} are not used here, we believe that the results
presented for the front instability and penetration modes are not strongly altered by this effect. Nevertheless, the effect of the stress singularity warrants further study.

In this work, density and viscosity ratios are set to unity. The characteristic velocity, $\U$, corresponds to $Q/A$, where $Q$ is the flow rate through the domain and $A$ is the cross-sectional area of the domain.
$\overline{D}$ is the mean cylinder diameter, which is sampled from a normal distribution.
We note that a normal distribution assumption may not be a representative of more
realistic geometries, but nonetheless we use it for a better control of the generated sizes that can be accommodated in our
finite computational domain. More representative distributions will be considered for future study. 
%
%
\section {Results and Discussions}
\label{sec:results}

The results are organized as follows. In Section \ref{sec:haines_jump_modes}, the primary mechanism of pore-filling, the Haines jump, as well as its two modes, are investigated. In section \ref{sec:displacement_regimes}, the different displacement regimes are characterized as functions of $\Ca$ and $\theta$. In Section \ref{sec:displacement_patterns}, the effects of the contact angle and capillary number on the shape of the propagation front of the invading phase are examined.

\begin{figure}[t]
	\centering
	\begin{overpic}[width=0.8\linewidth]{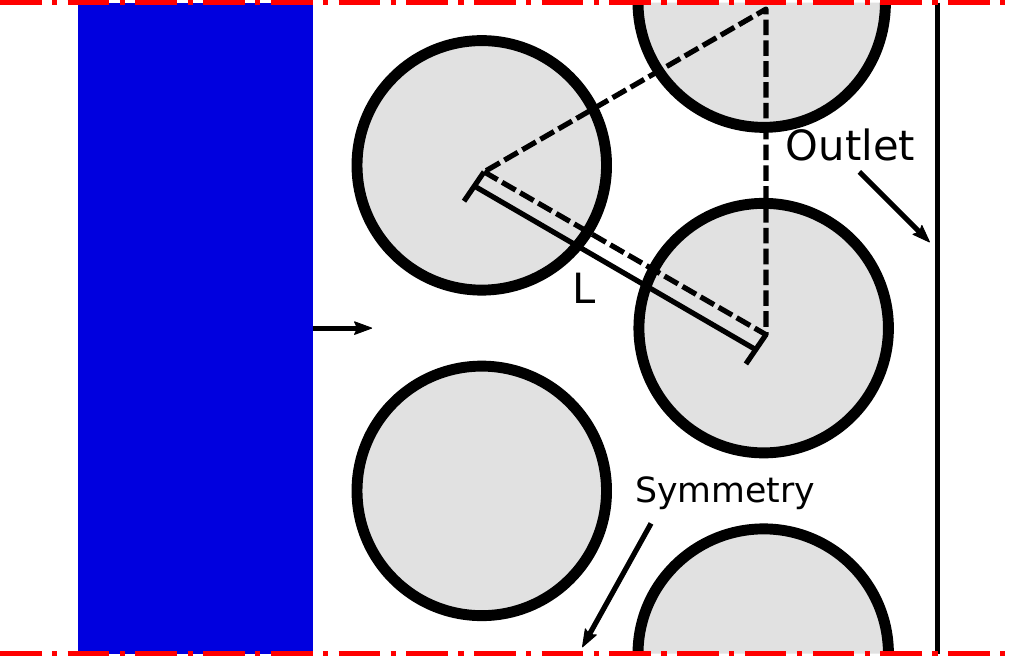}
		\put(31,28){$\vb{U}$}
	\end{overpic}
	\caption{Domain set-up and the corresponding boundary conditions for the Haines jump simulations.}
	\label{fig:SimSetup}
\end{figure}

\subsection{Haines Jump Modes}
\label{sec:haines_jump_modes}

\begin{figure}[t]
	\centering
	\begin{tabular}{ccc}
		\centering
		\includegraphics[width=0.32\linewidth]{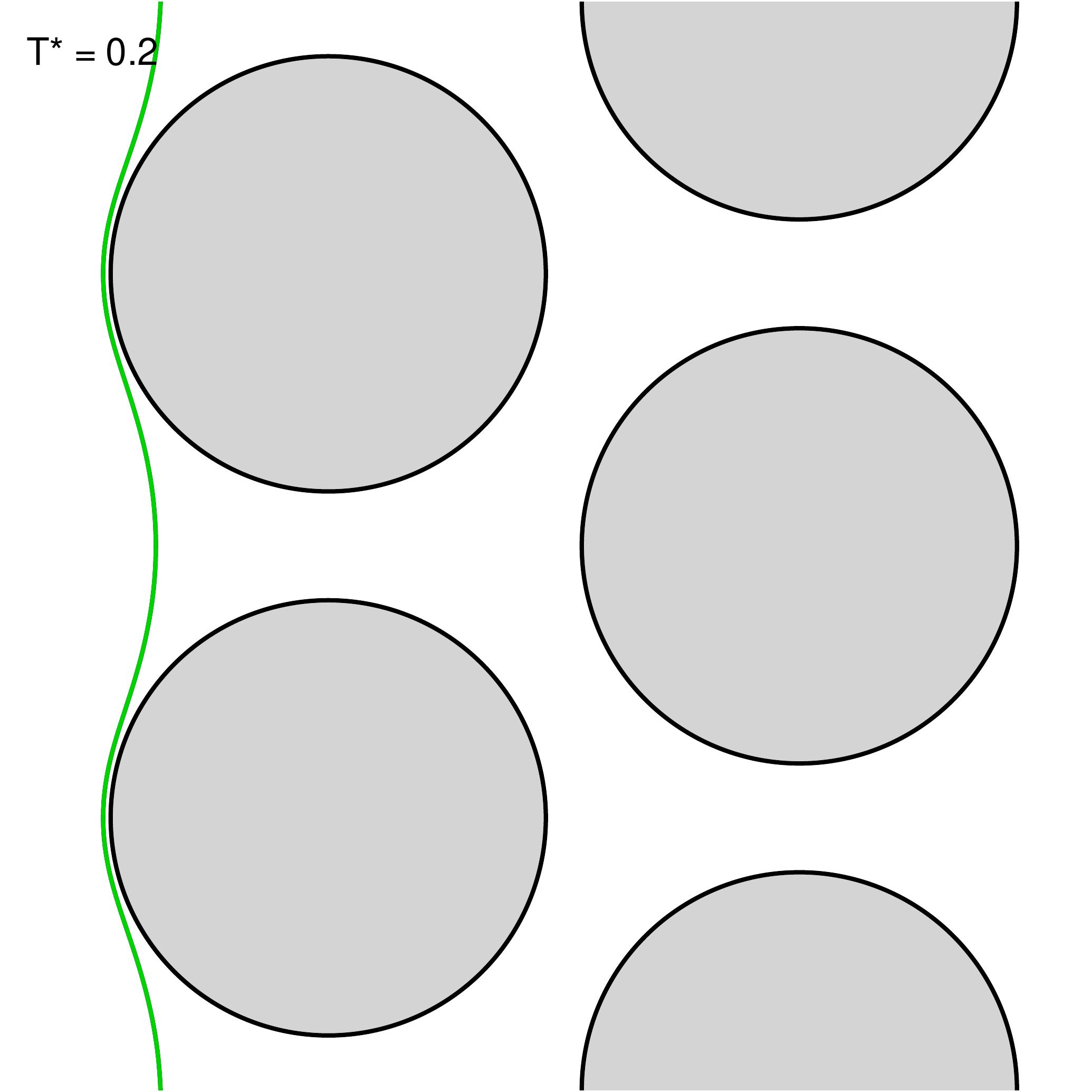}&
		\includegraphics[width=0.32\linewidth]{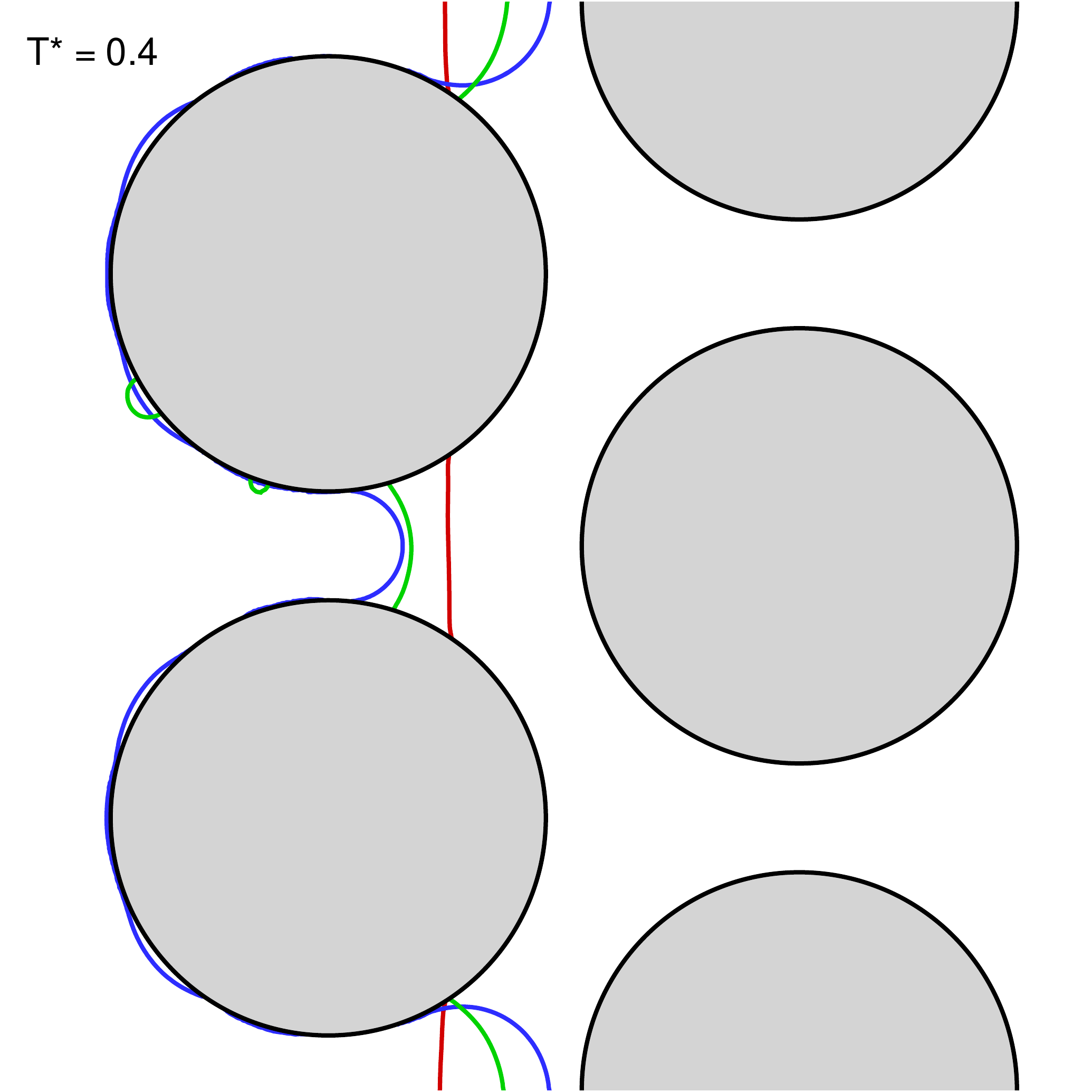}&
		\includegraphics[width=0.32\linewidth]{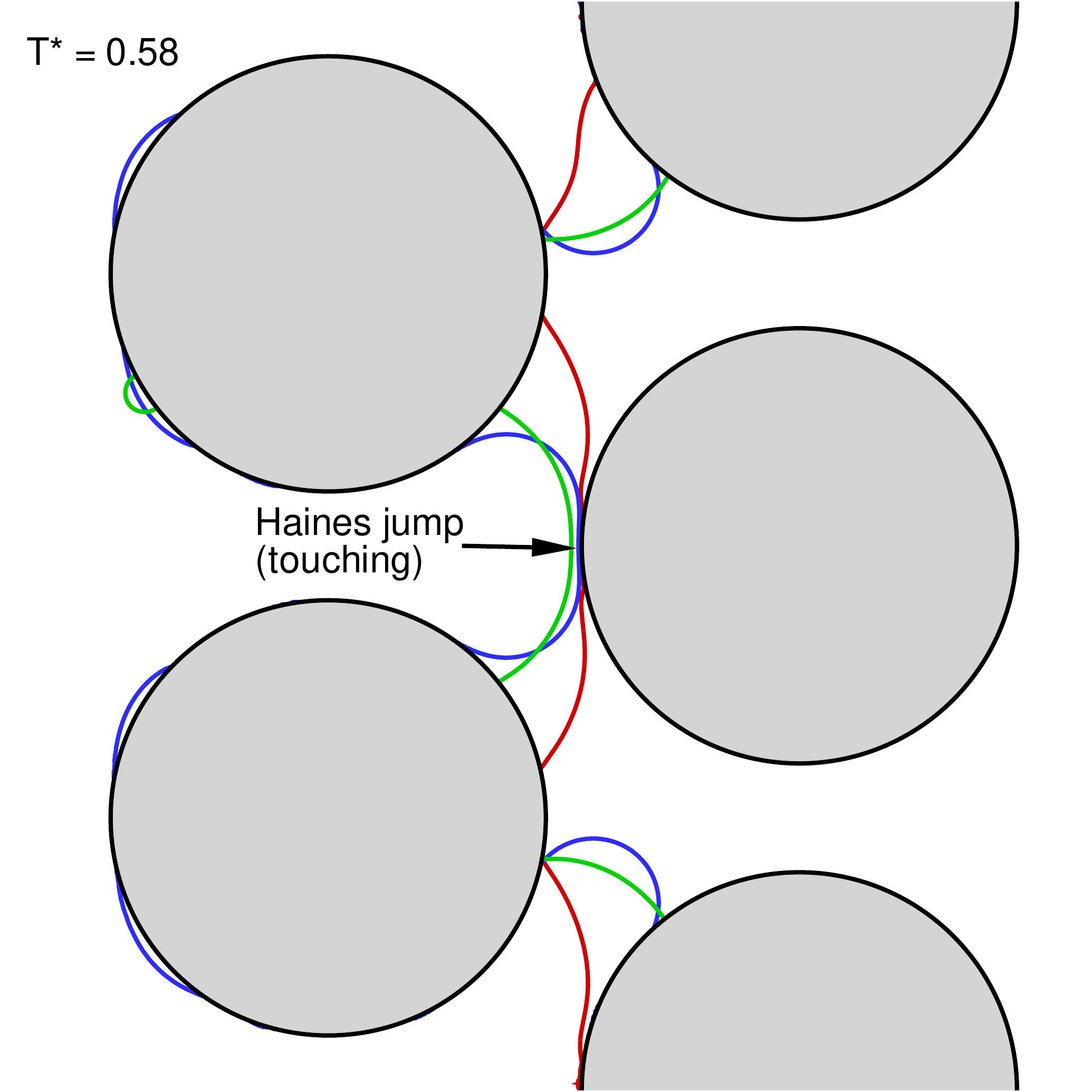}\\
		(a)&(b)&(c)\\
		\includegraphics[width=0.32\linewidth]{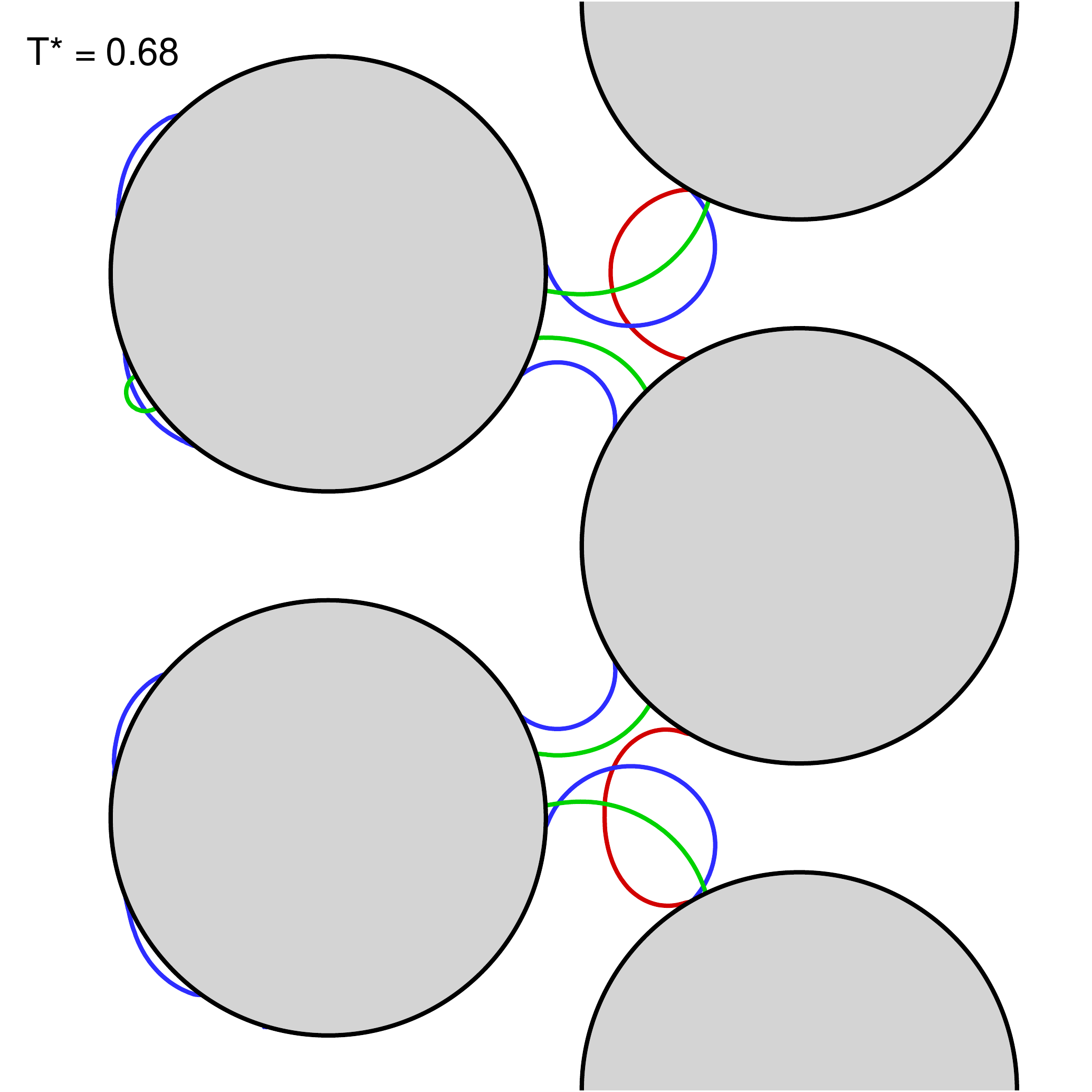}&
		\includegraphics[width=0.32\linewidth]{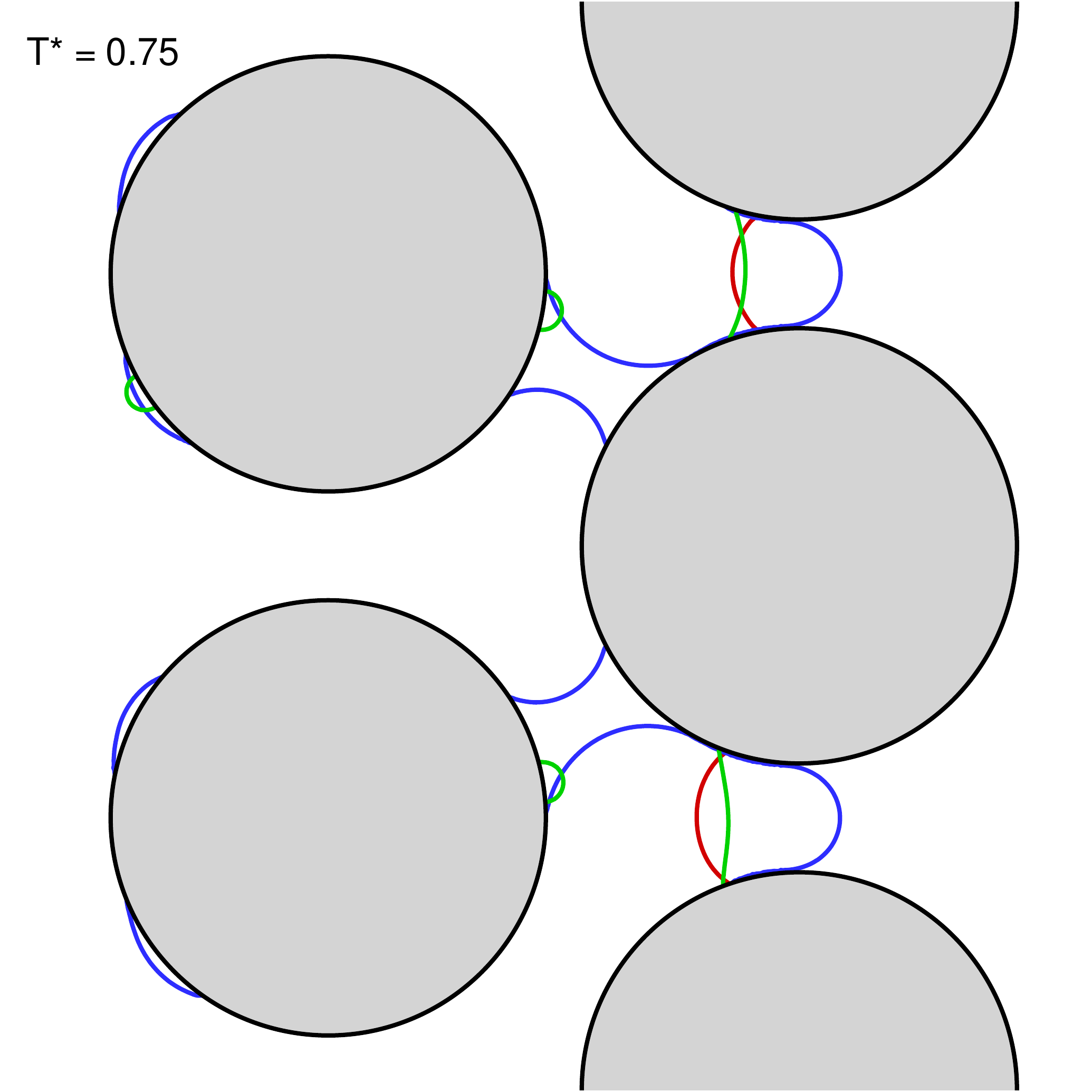}&
		\includegraphics[width=0.32\linewidth]{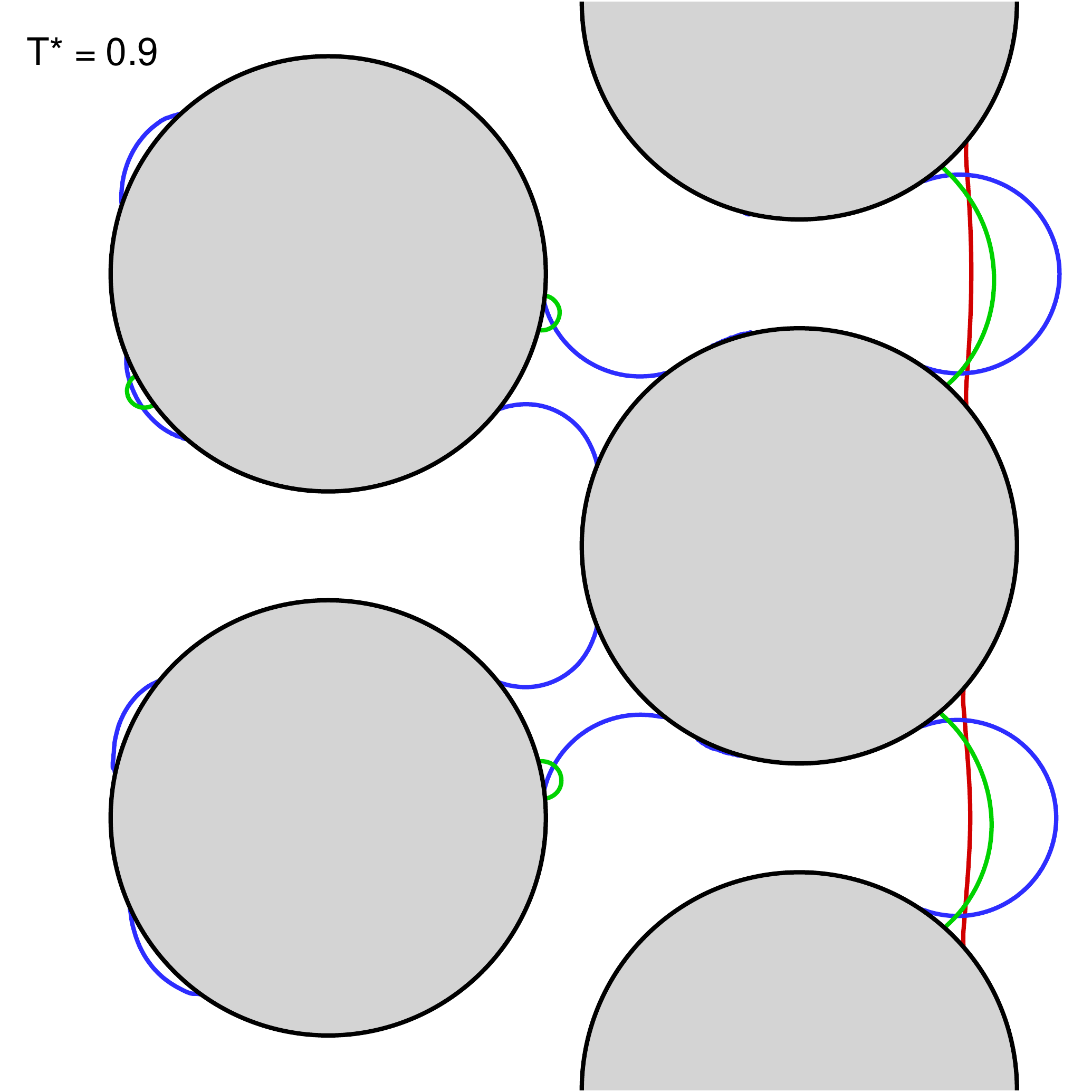}\\
		(d)&(e)&(f)\\
	\end{tabular}
	\caption{Simulation results displaying the Haines jump mechanism for $\theta=\ang{30}$ (\protect\redline), $\theta=\ang{90}$ (\protect\greenline), and $\theta=\ang{150}$ (\protect\blueline) for $\Ca=1\times10^{-3}$;
		$T^*$ = 0.2 (a), 0.4 (b), 0.58 (c), 0.68 (d), 0.75 (e), and 0.9 (f), where the non-dimensional time is $T^*=\overline{D}/\U$.}
	\vspace{-0.1in}
	\label{fig:haines_jump_Ca1e-3}
\end{figure}

\begin{figure}[t]
	\centering
	\begin{tabular}{ccc}
		\centering
		\includegraphics[width=0.32\linewidth]{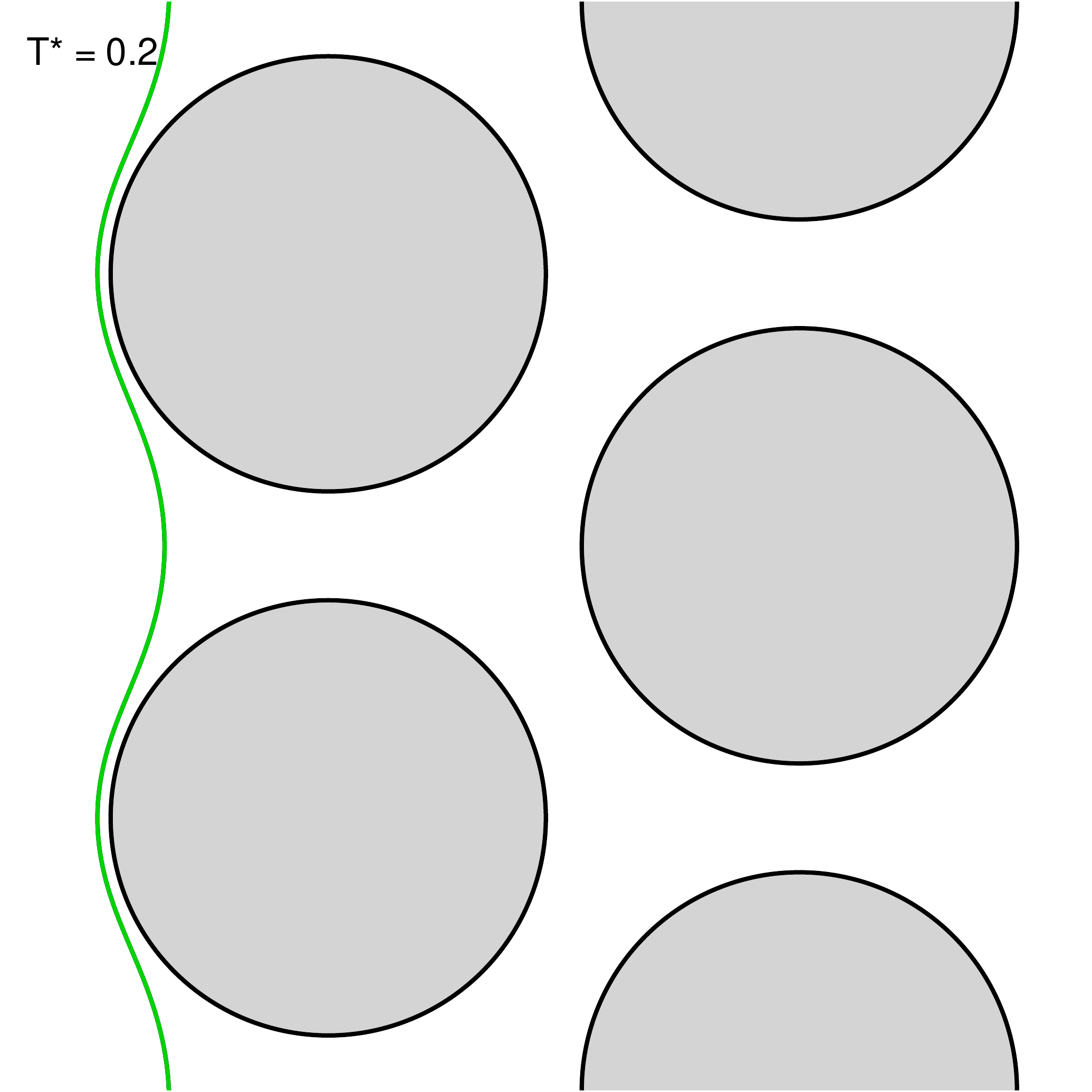}&
		\includegraphics[width=0.32\linewidth]{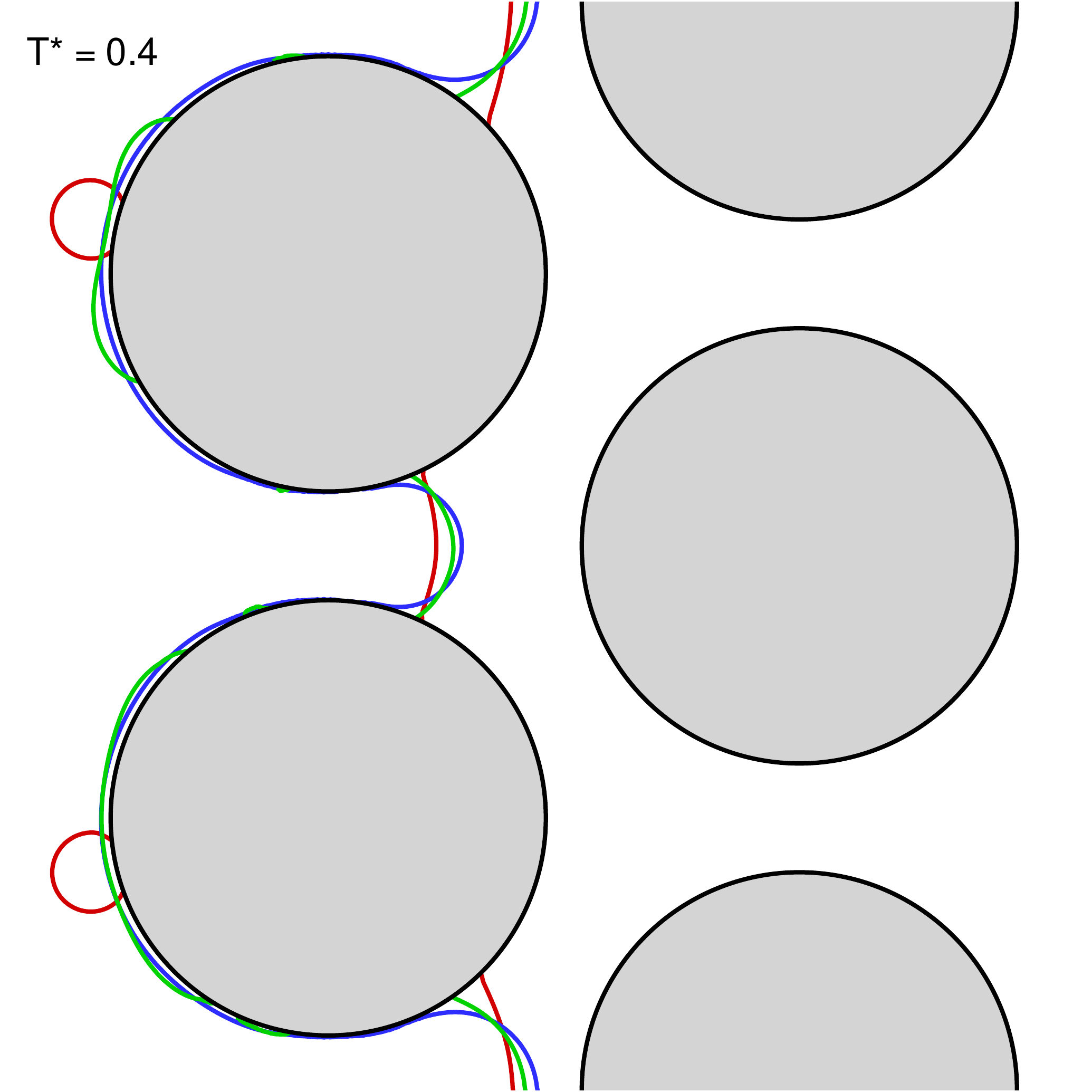}&
		\includegraphics[width=0.32\linewidth]{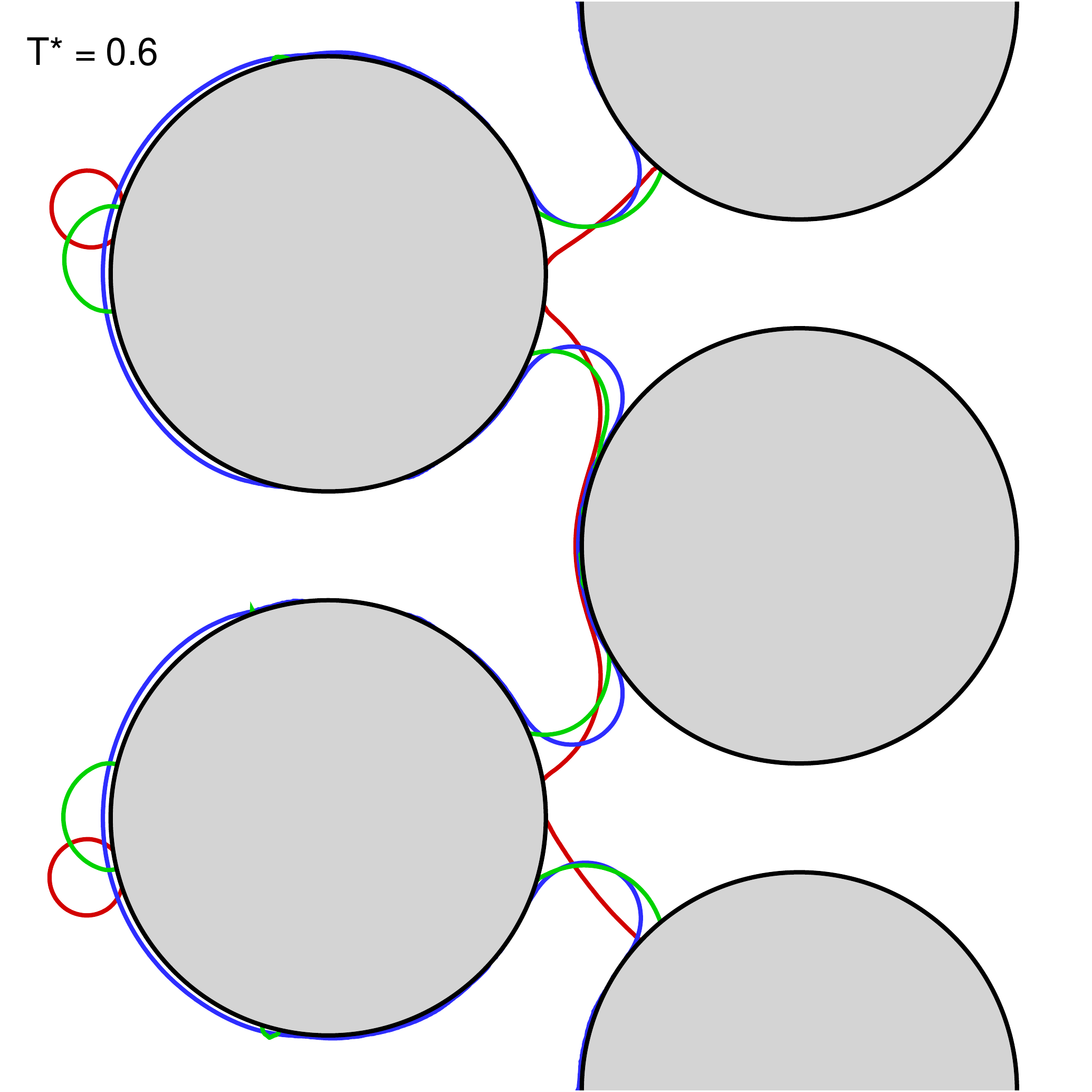}\\
		(a)&(b)&(c)\\
		\includegraphics[width=0.32\linewidth]{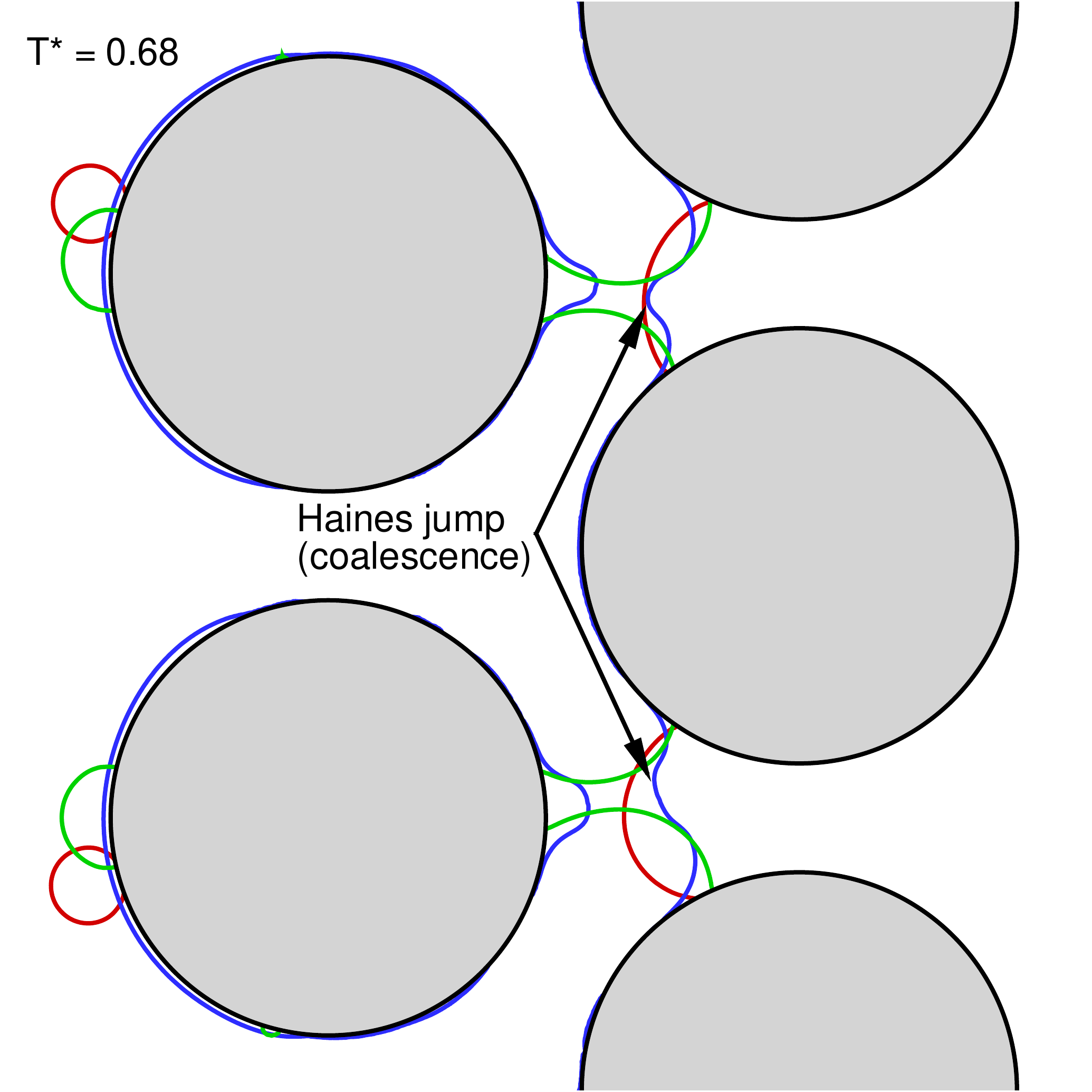}&
		\includegraphics[width=0.32\linewidth]{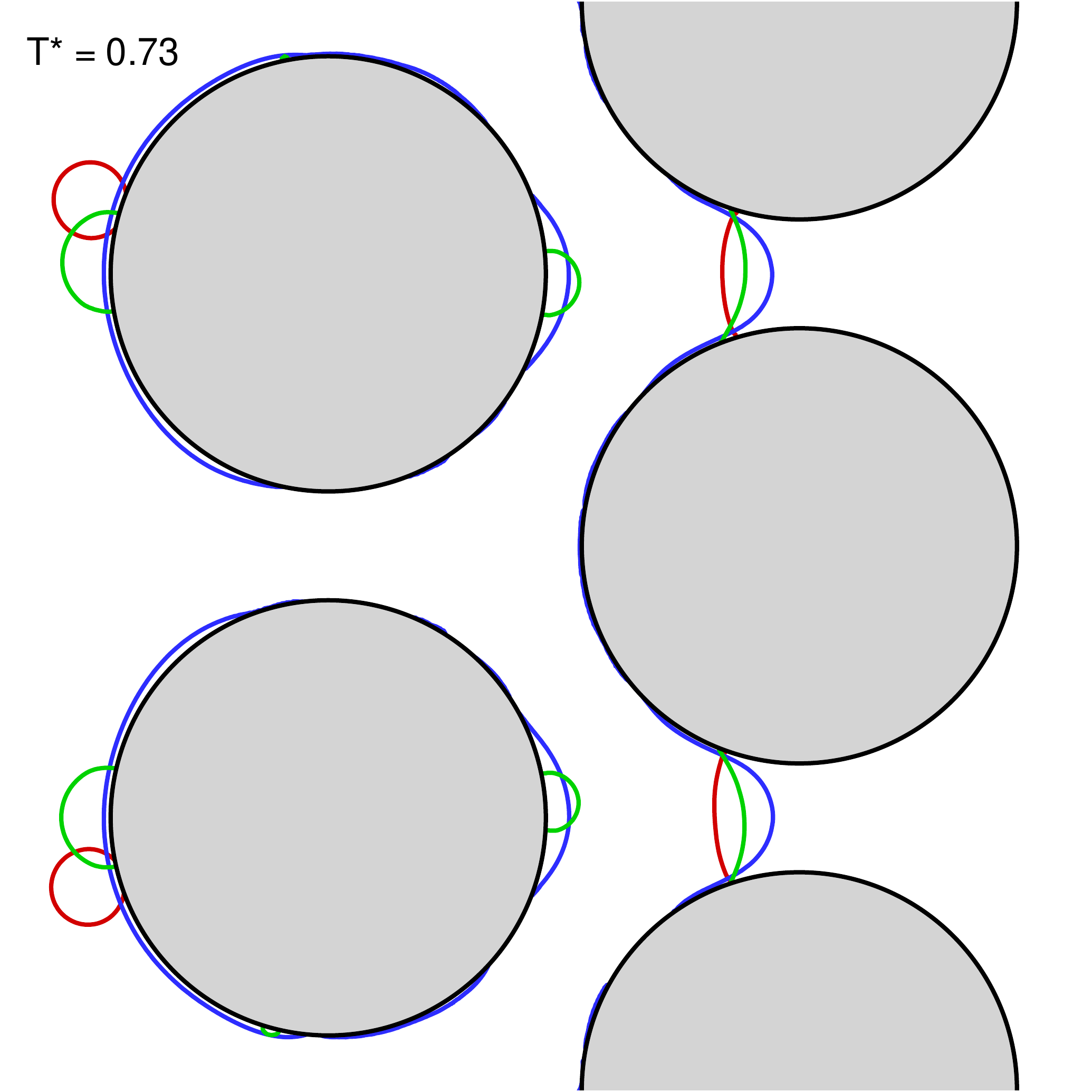}&
		\includegraphics[width=0.32\linewidth]{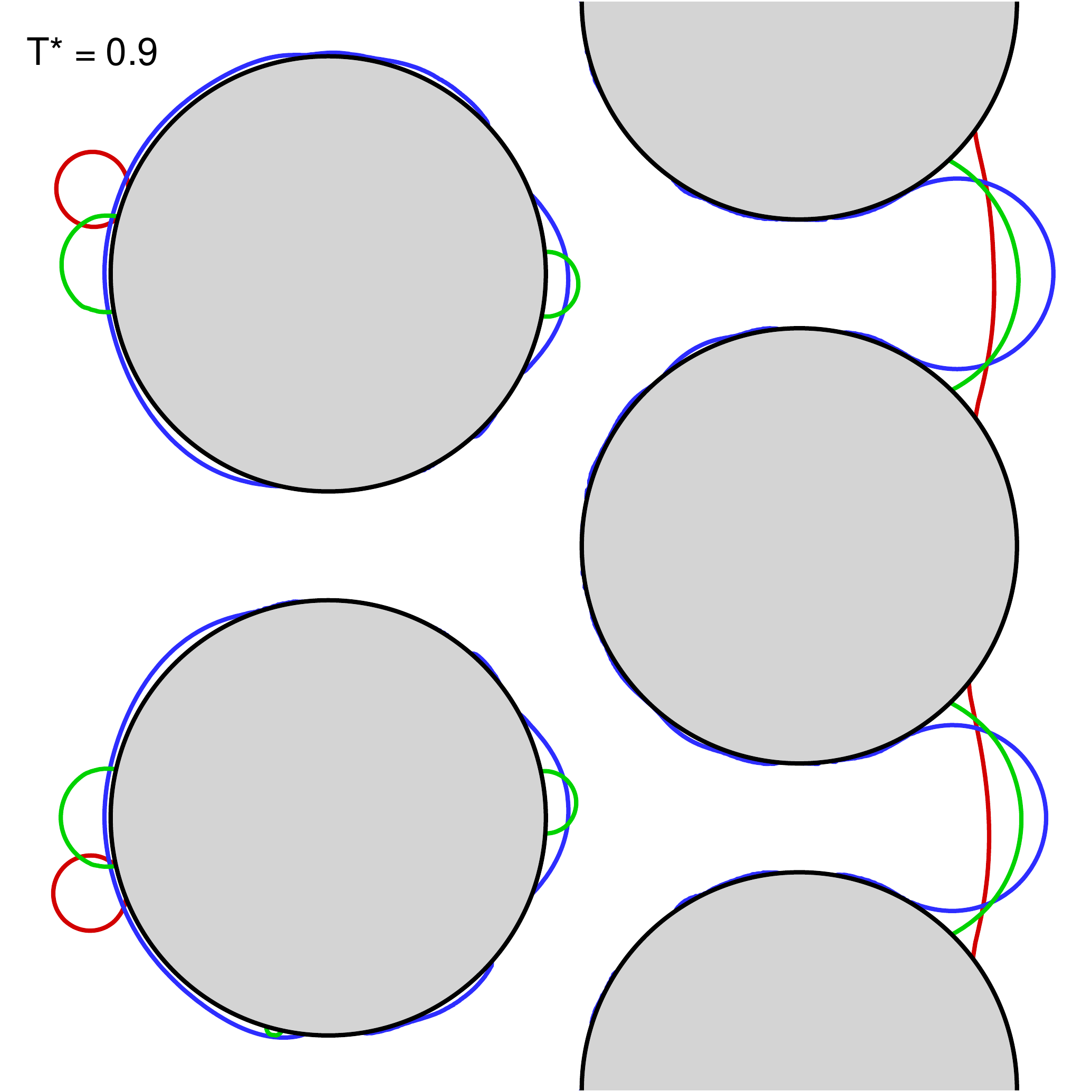}\\
		(d)&(e)&(f)\\
	\end{tabular}
	\caption{Simulation results displaying the Haines jump mechanism for $\theta=\ang{30}$ (\protect\redline), $\theta=\ang{90}$ (\protect\greenline), and $\theta=\ang{150}$ (\protect\blueline) for $\Ca=2.5\times10^{-3}$;
		$T^*$ = 0.2 (a), 0.4 (b), 0.6 (c), 0.68 (d), 0.73 (e), and 0.9 (f).}
	\vspace{-0.1in}
	\label{fig:haines_jump_Ca2.5e-3}
\end{figure}

\begin{figure}[t]
	\centering
	\begin{tabular}{ccc}
		\centering
		\includegraphics[width=0.32\linewidth]{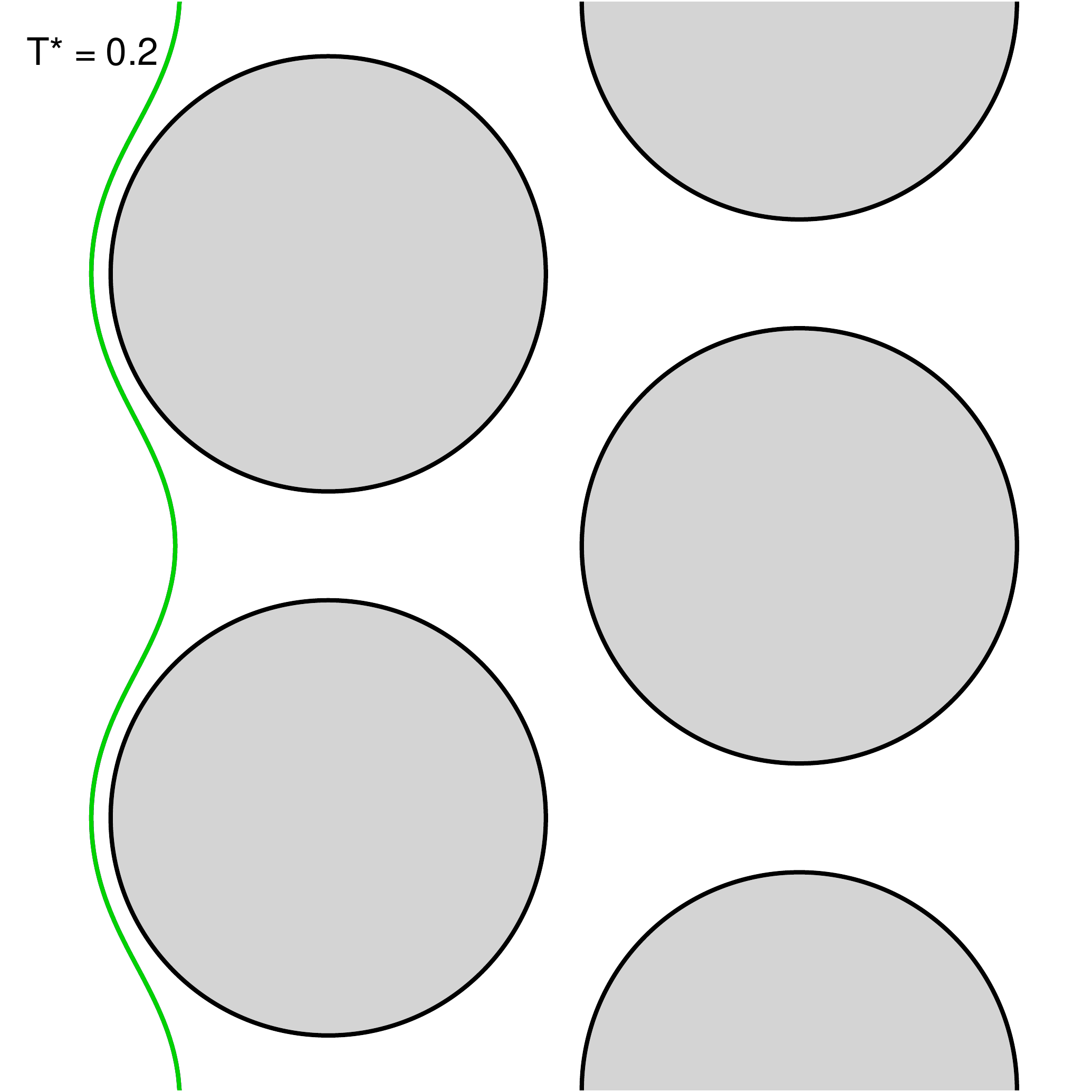}&
		\includegraphics[width=0.32\linewidth]{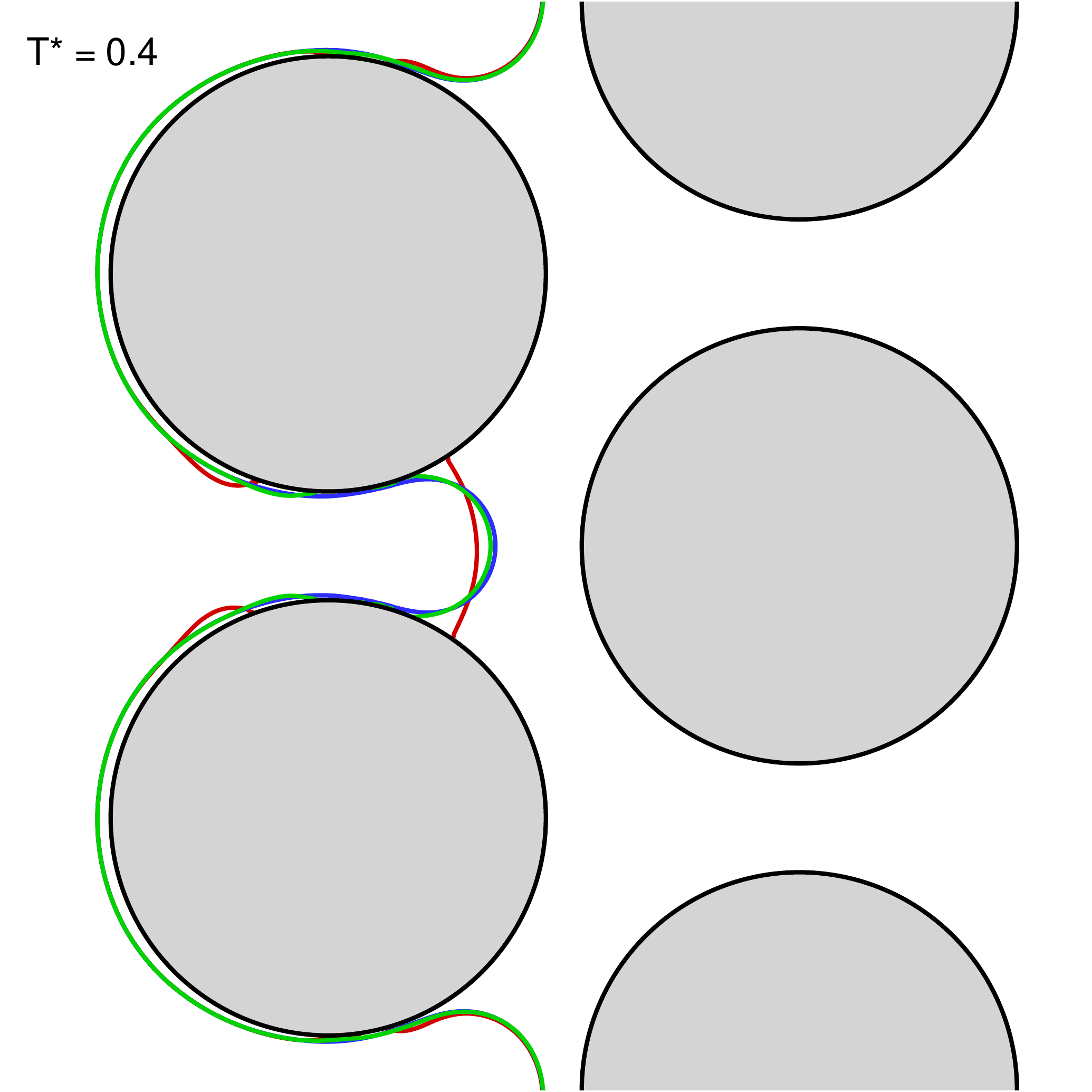}&
		\includegraphics[width=0.32\linewidth]{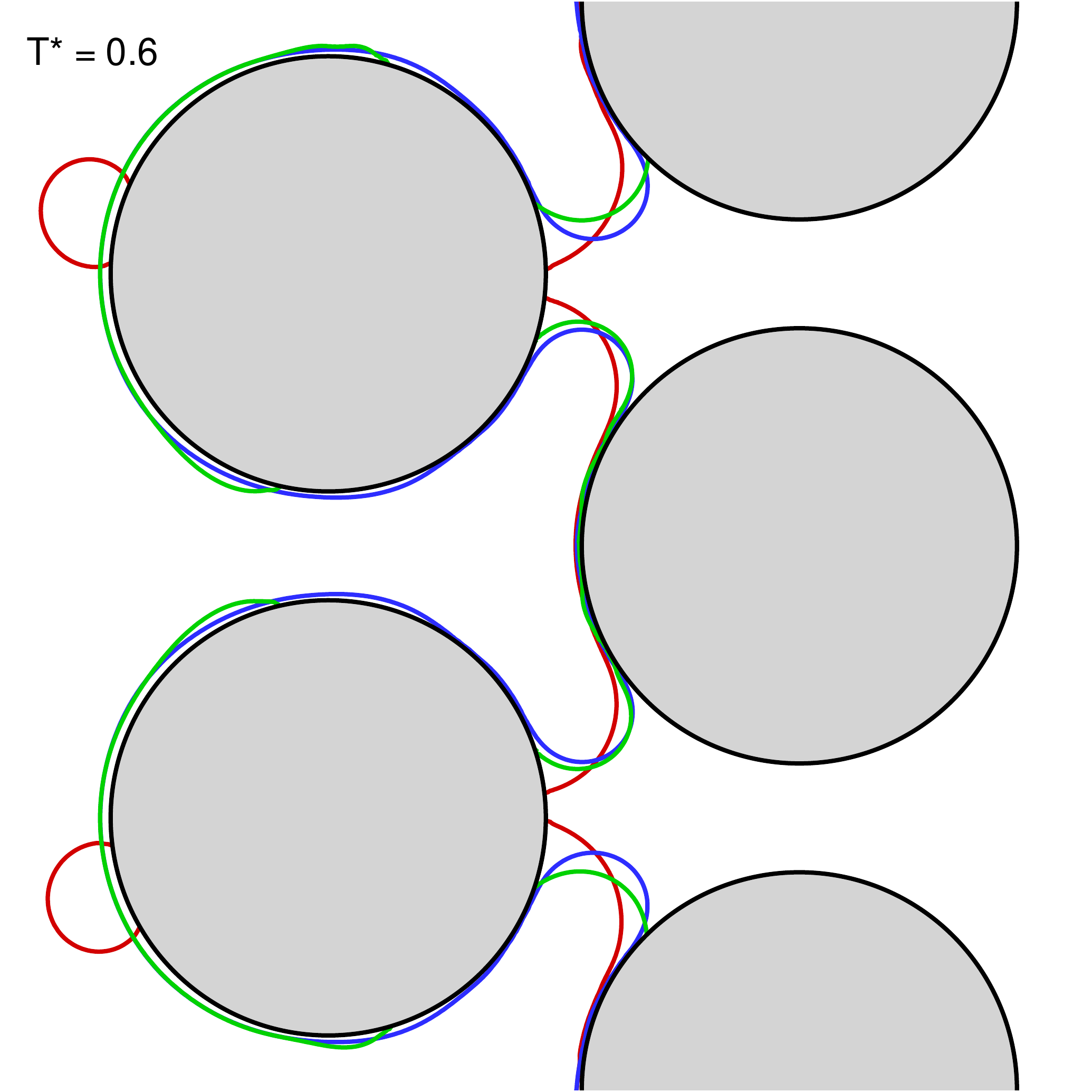}\\
		(a)&(b)&(c)\\
		\includegraphics[width=0.32\linewidth]{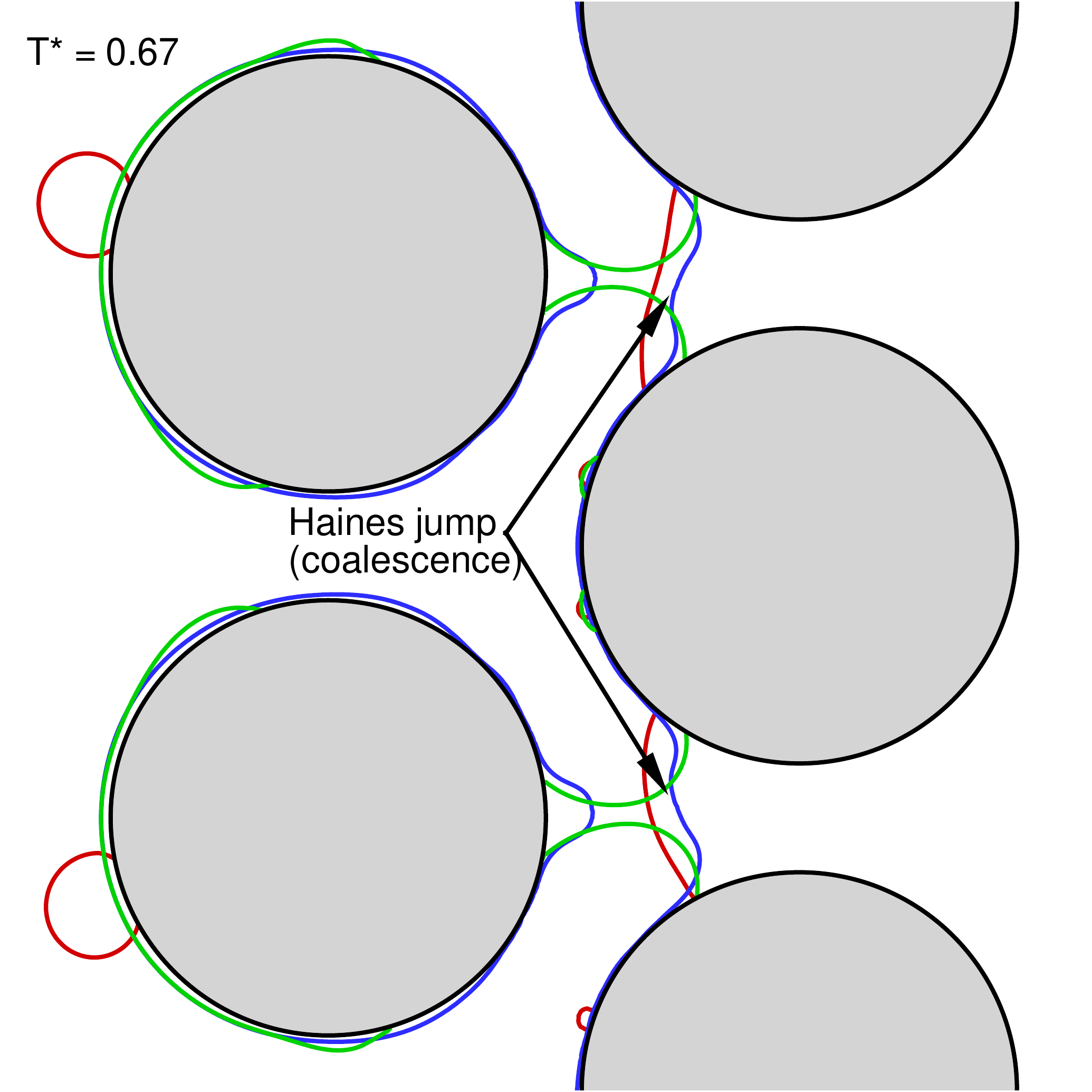}&
		\includegraphics[width=0.32\linewidth]{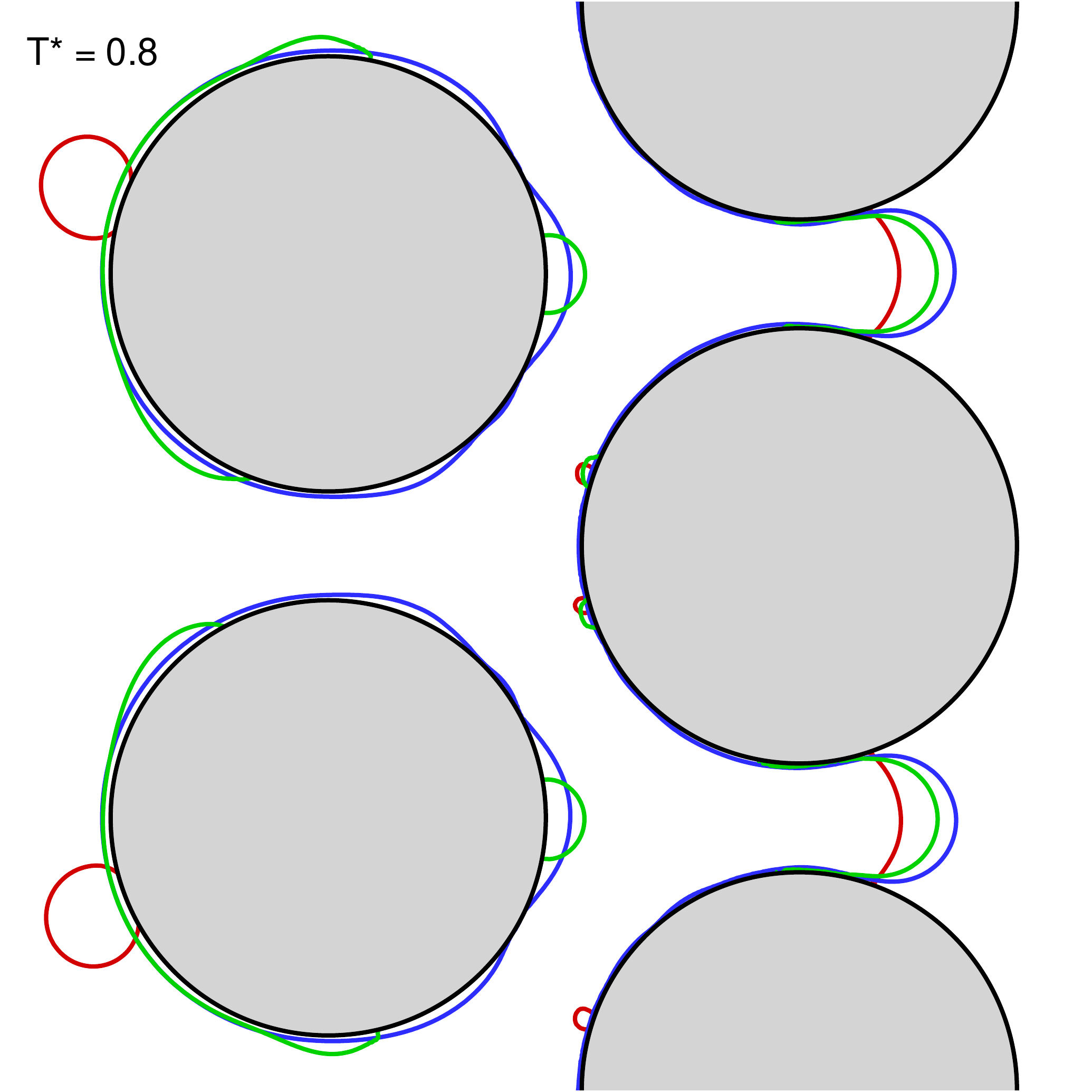}&
		\includegraphics[width=0.32\linewidth]{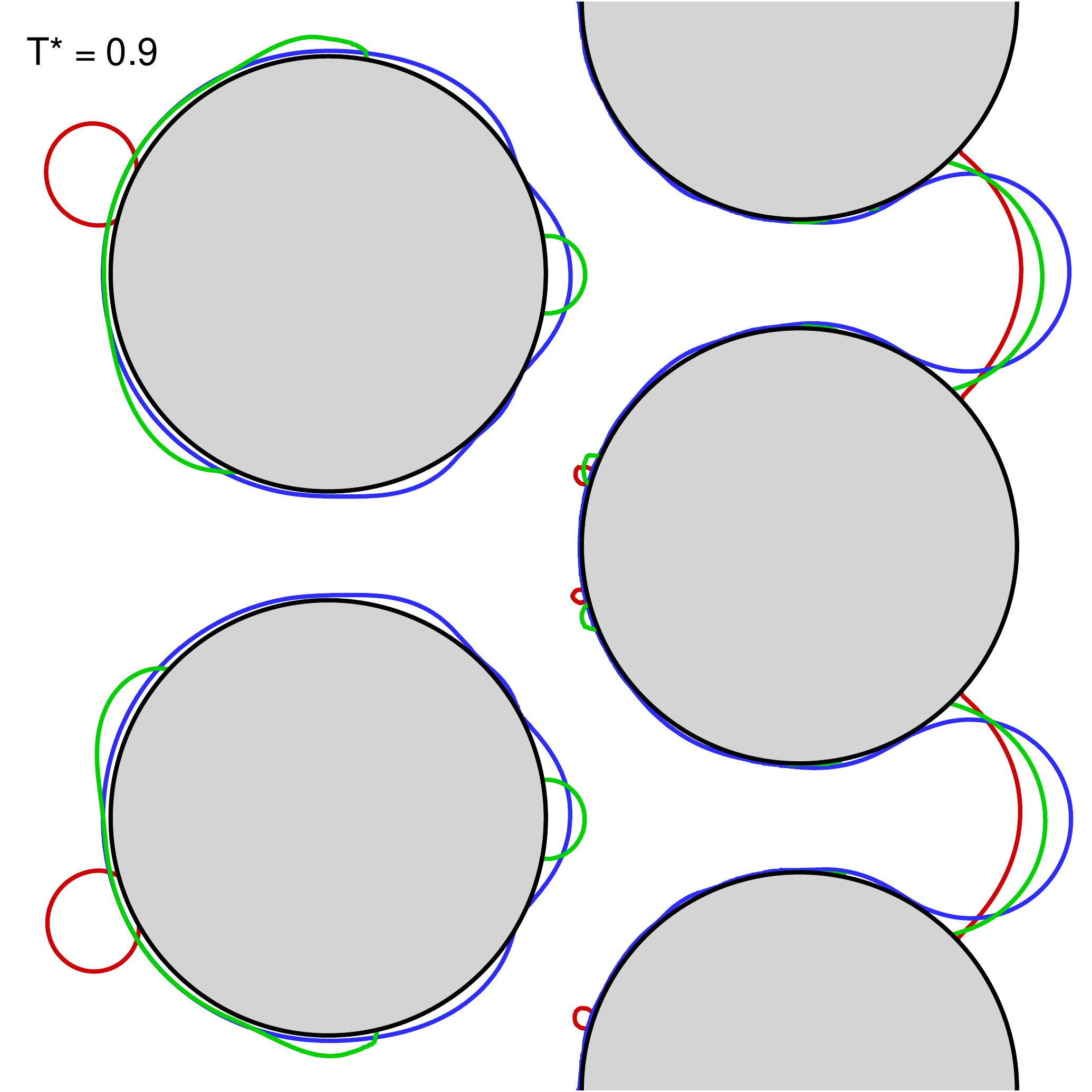}\\
		(d)&(e)&(f)\\
	\end{tabular}
	\caption{Simulation results displaying the Haines jump mechanism for $\theta=\ang{30}$ (\protect\redline), $\theta=\ang{90}$ (\protect\greenline), and $\theta=\ang{150}$ (\protect\blueline) for $\Ca=5\times10^{-3}$;
		$T^*$=0.2 (a), 0.4 (b), 0.6 (c), 0.67 (d), 0.8 (e), and 0.9 (f).}
	\vspace{-0.1in}
	\label{fig:haines_jump_Ca5e-3}
\end{figure}

A simplified setup, depicted in Fig.\ \ref{fig:SimSetup}, is used in order to isolate the primary mechanisms of pore-filling. 
Three different $\Ca$ are compared, and the results are illustrated in Figs.\ \ref{fig:haines_jump_Ca1e-3} to \ref{fig:haines_jump_Ca5e-3}. Singh \etal \cite{Singh2017} identified two different geometry-dependent Haines jump modes for cylinders and spheres. In 2D, these modes effectively correspond to the `touching' and `coalescence' modes described in \cite{Cieplak1988,Cieplak1990}. We
use the constructed geometry in Fig.\ \ref{fig:SimSetup} to qualitatively demonstrate both modes. We find that
both modes of Haines jump will only occur at certain values of $\Ca$. 
As the interfaces corresponding to $\theta=\ang{30}$, $\theta=\ang{90}$, and $\theta=\ang{150}$ pass between the two left-most cylinders, they begin to `bulge' until they make contact with the center cylinder (Fig.\ \ref{fig:haines_jump_Ca1e-3}c). This formation of a new contact line on the center cylinder initiates a rapid reconfiguration of the interface due to unbalanced surface tension forces. This represents the first mode of the Haines jump. As $\Ca$ is increased, and as the interfaces approach the throats formed by the right-most cylinders, they merge with a secondary interface passing around a cylinder (Figs.\ \ref{fig:haines_jump_Ca2.5e-3}d and \ref{fig:haines_jump_Ca5e-3}d). This causes a rapid release of the capillary pressure gradient acting against the flow, which causes the invading fluid to rapidly fill the pore and move through the throat. This corresponds to the second mode of the Haines jump. 
The effect of the geometry, injection rate and capillary number on the Haines jumps were studied in detail by Armstrong \etal \cite{Armstrong2015}, showing that the occurrence of a Haines jump could be predicted from simple geometrical arguments. Here we also show that at low invading phase contact angles, and when the media is not densely packed, it is then possible that neither mode of Haines jump will occur. For the first mode, the contact lines on the adjacent sides of an obstacle will simply merge, and the protruding of the interface from the throat necessary for the second mode to occur will not be observed. In this scenario, complete displacement of the defending phase will be observed. When a Haines jump occurs, a pathway forms for the fluid to travel, in which no resistance from capillary pressure is present. As a result, pockets of defending fluid may remain in place, since the invading fluid will simply flow around them through sites in which Haines jumps have occurred.

As seen in Fig.\ \ref{fig:haines_jump_Ca5e-3}, a viscous boundary layer forms initially, preventing a contact line  to form on the middle cylinder before the contact lines traversing the surface of the left-most cylinders merge. At low $\Ca$, as seen in Fig.\ \ref{fig:haines_jump_Ca1e-3}, the driving pressure may be too small to overcome the capillary forces at the throat, and the interface stalls, effectively blocking flow through the throat. At higher $\Ca$ (see Figs.\ \ref{fig:haines_jump_Ca2.5e-3} and \ref{fig:haines_jump_Ca5e-3}), the liquid film trapped along the solid boundary collapses into a droplet upon the formation of contact lines for lower invading phase contact angles. Decreasing $\Ca$ results in a more stable contact line, with a $\theta$ that corresponds more closely to the prescribed equilibrium value. 

We also note that the resolution needs to be sufficient to appropriately resolve the small features noticed above. 
Our developed numerical schemes provide a robust and high accuracy solution to capture these small structures within the pores. 
References \citep{OBrien2018} and \citep{OBrien2019} detail the numerical implementation in this work that allows to accurately capture such features. 
However, we have not investigated the required resolution to resolve these small length scales,
as we do not believe that the phenomena of interest here, which is the way that the large length scale front invades the pore under different contact angles,
is influenced significantly by this kind of small scale structure.

\subsection{Displacement Regimes}
\label{sec:displacement_regimes}

\begin{center}
	\begin{figure}[t]
		\includegraphics[width=1.15\textwidth,trim=10mm 0 0 0]{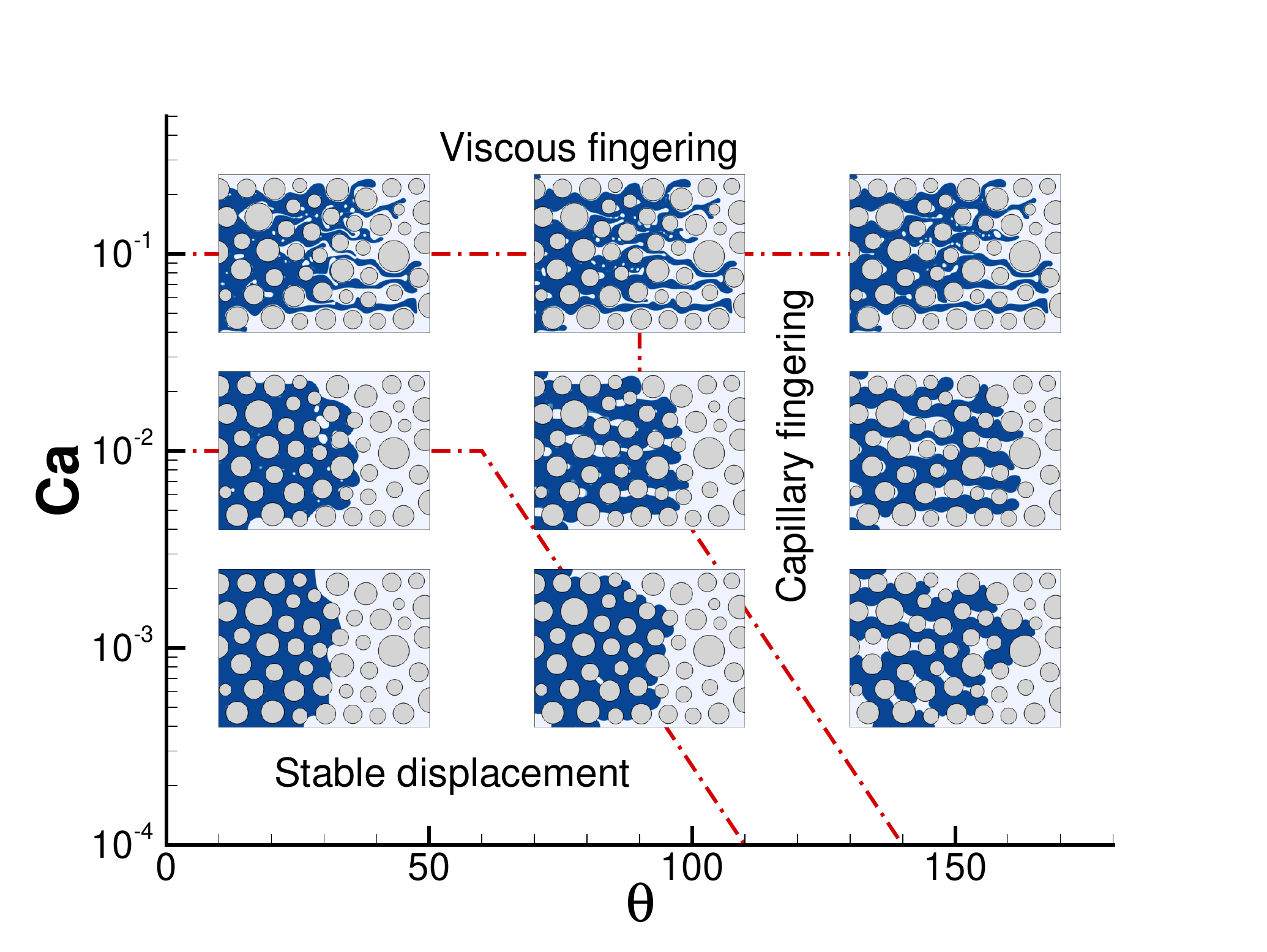}
		\caption{Displacement patterns at varying $\Ca$ and $\theta$. Above a certain $\Ca$ threshold only viscous fingering is visible for the parameters studied in this work.}
		\label{fig:displacement_patterns}
		\vspace{-0.1in}
	\end{figure}
\end{center}

The primary motivation behind this work is to investigate the effect that varying invading phase contact angle has on the 
fraction  of  the defending fluid that has been displaced from the domain in a porous media model.
In order to investigate this, a random array of randomly-sized cylinders was generated, with different prescribed contact angles ranging from $\theta = \ang{30}$ to $\theta=\ang{150}$. The secondary phase is injected into the cylinder array at a constant rate, at $\rRe=5$, where the characteristic length is defined as the mean cylinder diameter. The displacement regimes include viscous fingering (high $\Ca$), capillary fingering (low $\Ca$, $\theta > \ang{90}$) and stable displacement (low $\Ca$, $\theta < \ang{90}$). The existence of these regimes has been discussed by previous authors \cite{Zhang2011,Ferrari2013,Mei2012,Zacharoudiou2018,Lenormand1990,Jung2016} for different independent parameters. Injection patterns as a function of $\Ca$ and $\theta$ are shown in Fig.\ \ref{fig:displacement_patterns}. Viscous fingering can be identified by the formation of long, thin fingers of the invading phase which often do not touch the obstacles as the relatively weaker surface tension results in the fingers simply following the velocity streamlines. Capillary fingers form contact lines with the obstacles, and are much rounder in shape due to the relatively strong surface tension forces present with high curvatures. In stable displacement, the front is much flatter and the interface much less distorted. Based on the observed data, transitions between the three known displacement regions can be identified, in a similar manner to how Zhang \etal \cite{Zhang2011} identified the existence of different displacement regions for a range of $\Ca$ and viscosity ratios. Fig.\ \ref{fig:displacement_patterns} depicts the transition between different regions in terms of $\Ca$ and $\theta$.

As seen in Fig.\ \ref{fig:displacement_patterns}, as the viscosity ratio is held constant at one, all cases of $\Ca > 0.1$ represent examples of viscous fingering, with little difference observed between media at different contact angles. As $\Ca$ drops and $\theta$ begins to increase, capillary fingering behavior is observed. Stable fronts are generally characterized by low $\Ca$ and low
invading phase contact angles. These regimes were not observed to change much in our numerical experiments when slightly varying $\Re$; nevertheless, for the applications we consider in this work, 
we keep $\Re$ small. However, investigating the  effects of varying $\Re$ on the results is of interest and since it is beyond the scope of this study, we leave 
it for future work.

\subsection{Displacement Patterns}
\label{sec:displacement_patterns}

\begin{figure}[t]
	\centering
	\begin{tabular}{ccc}
		\includegraphics[width=.5\textwidth]{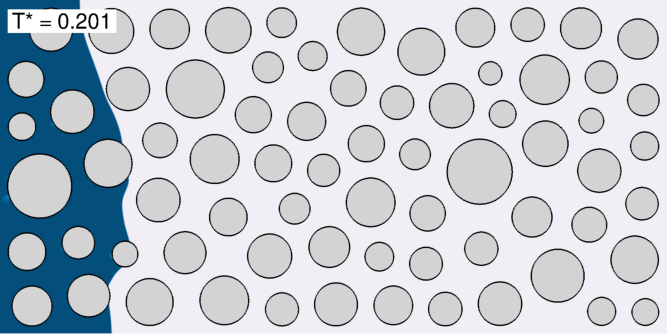}&
		\includegraphics[width=.5\textwidth]{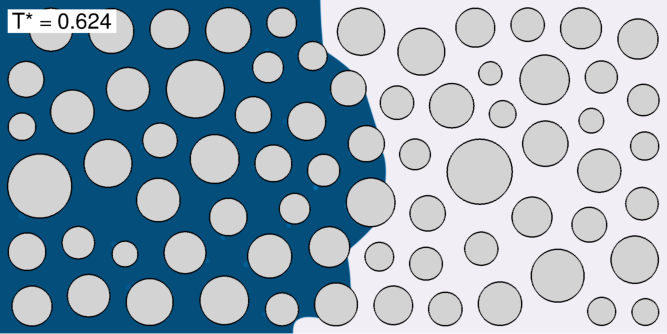}\\
		\includegraphics[width=.5\textwidth]{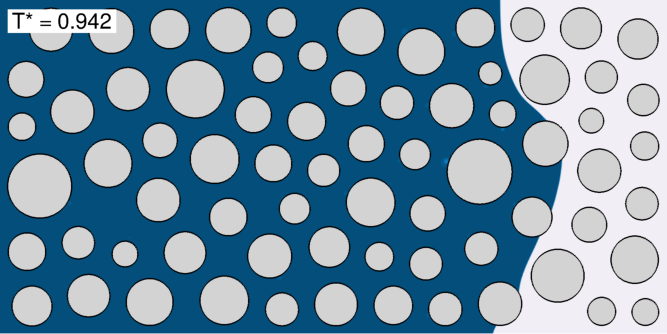}&
		\includegraphics[width=.5\textwidth]{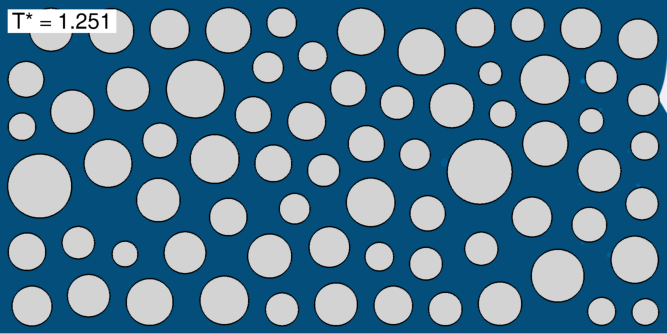}\\
	\end{tabular}
	\caption{Injection simulation for $\theta=\ang{30}$, with $\rRe=5$ and $\Ca=1\times10^{-3}$.}
	\label{fig:injection_th30}
	\vspace{-0.1in}
\end{figure}

\begin{figure}[!h]
	\centering
	\begin{tabular}{ccc}
		\includegraphics[width=.5\textwidth]{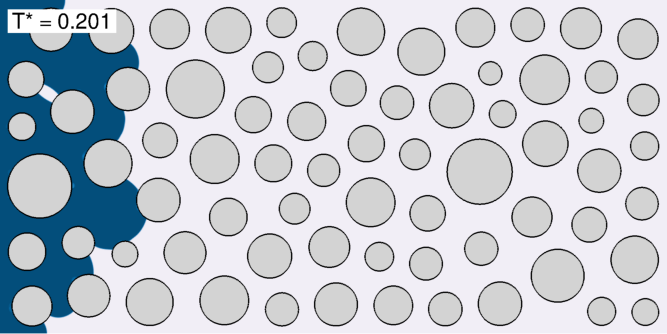}&
		\includegraphics[width=.5\textwidth]{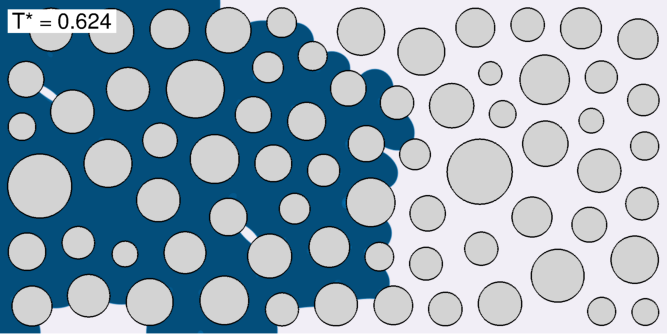}\\
		\includegraphics[width=.5\textwidth]{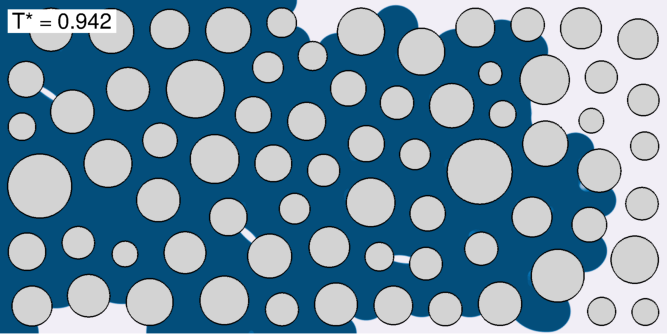}&
		\includegraphics[width=.5\textwidth]{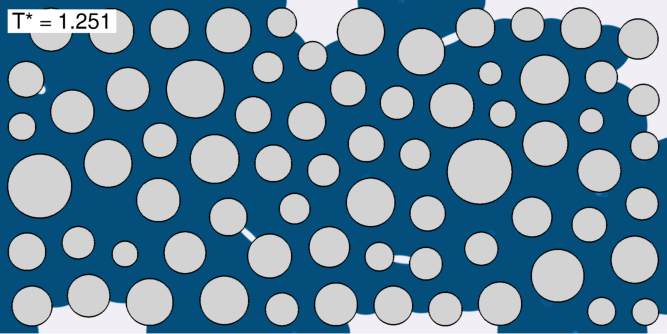}\\
	\end{tabular}
	\caption{Injection simulation for $\theta=\ang{90}$, with $\rRe=5$ and $\Ca=1\times10^{-3}$.}
	\label{fig:injection_th90}
	\vspace{-0.1in}
\end{figure}

\begin{figure}[!h]
	\centering
	\begin{tabular}{ccc}
		\includegraphics[width=.5\textwidth]{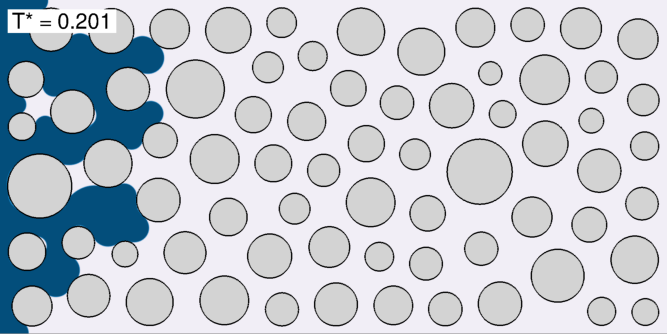}&
		\includegraphics[width=.5\textwidth]{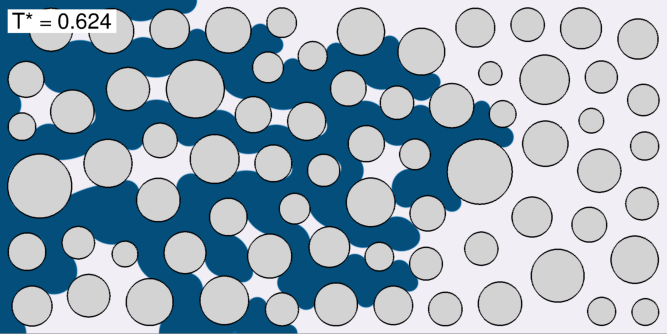}\\
		\includegraphics[width=.5\textwidth]{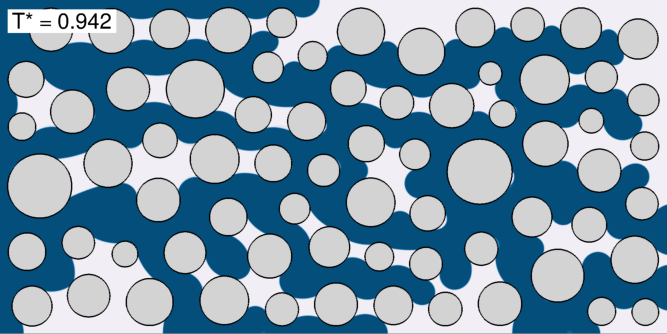}&
		\includegraphics[width=.5\textwidth]{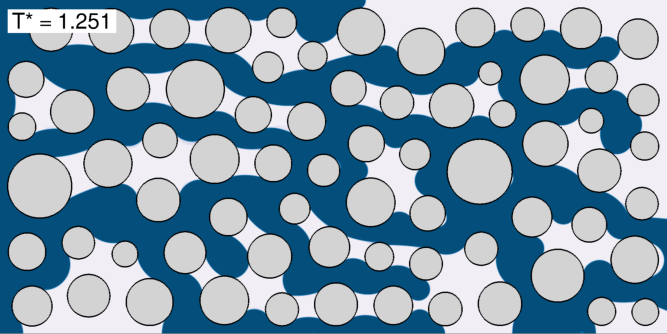}\\
	\end{tabular}
	\caption{Injection simulation for $\theta=\ang{150}$, with $\rRe=5$ and $\Ca=1\times10^{-3}$.}
	\label{fig:injection_th150}
	\vspace{-0.1in}
\end{figure}

Here we demonstrate the progression of the front through the porous media described in the previous section, for $\Ca=10^{-3}$.  Figs.\ \ref{fig:injection_th30}, \ref{fig:injection_th90} and \ref{fig:injection_th150} depict snapshots of the fluid injection into the porous media for contact angles of $\theta=\ang{30}$, $\theta=\ang{90}$ and $\theta=\ang{150}$ respectively.

\begin{figure}[th]
	\begin{tabular}{ccc}
		\includegraphics[width=0.32\linewidth,trim=20mm 0 0 0]{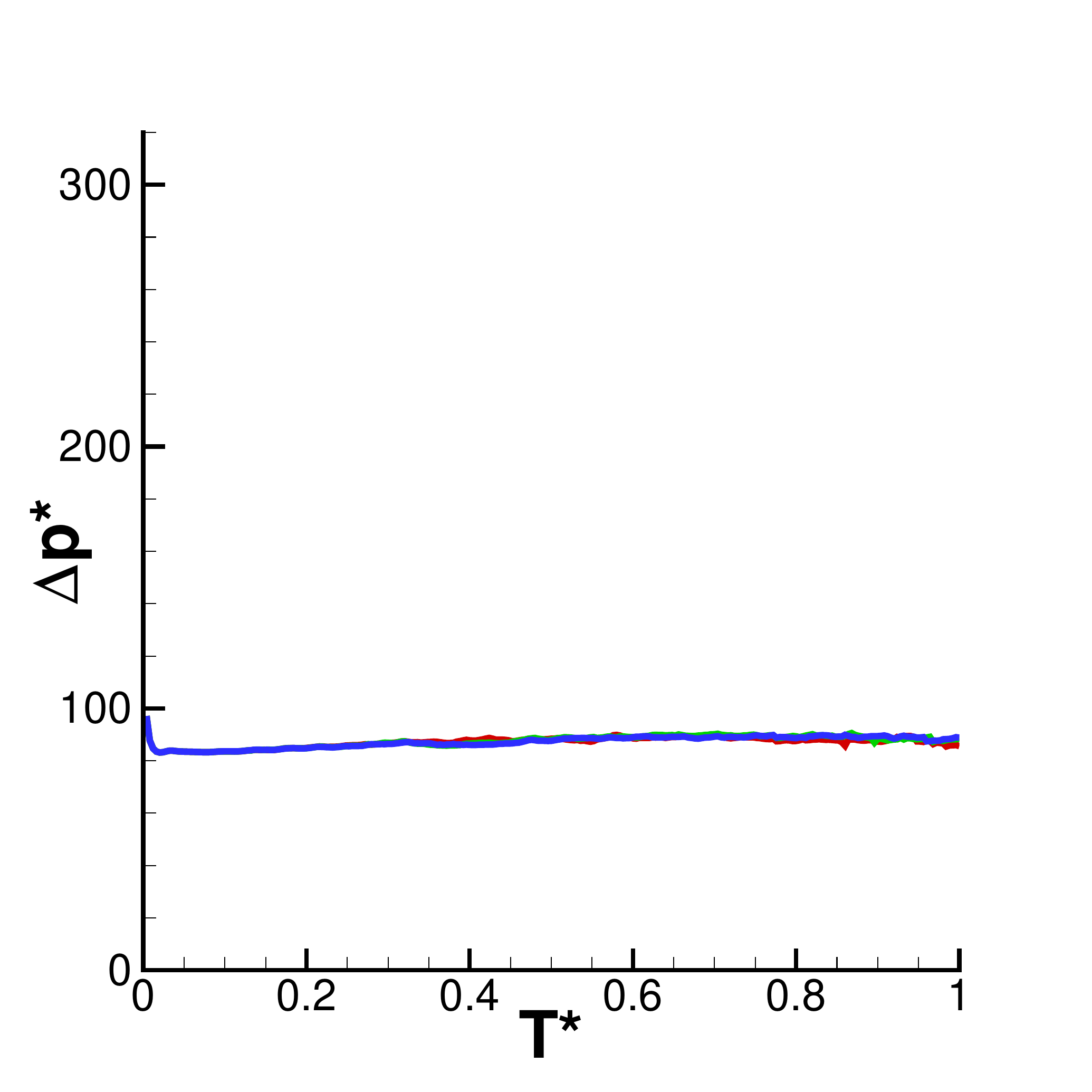}&
		\includegraphics[width=0.32\linewidth,trim=20mm 0 0 0]{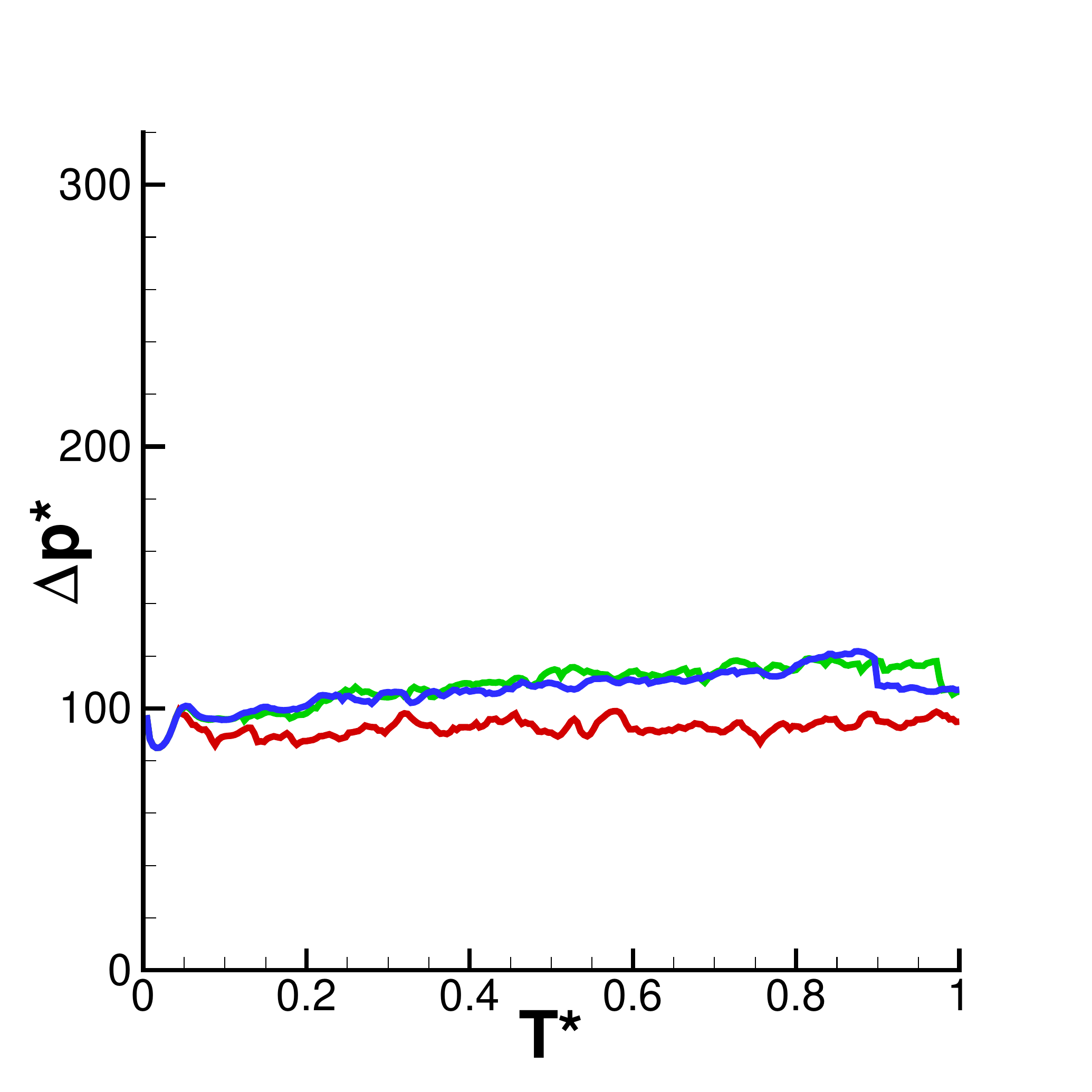}&
		\includegraphics[width=0.32\linewidth,trim=20mm 0 0 0]{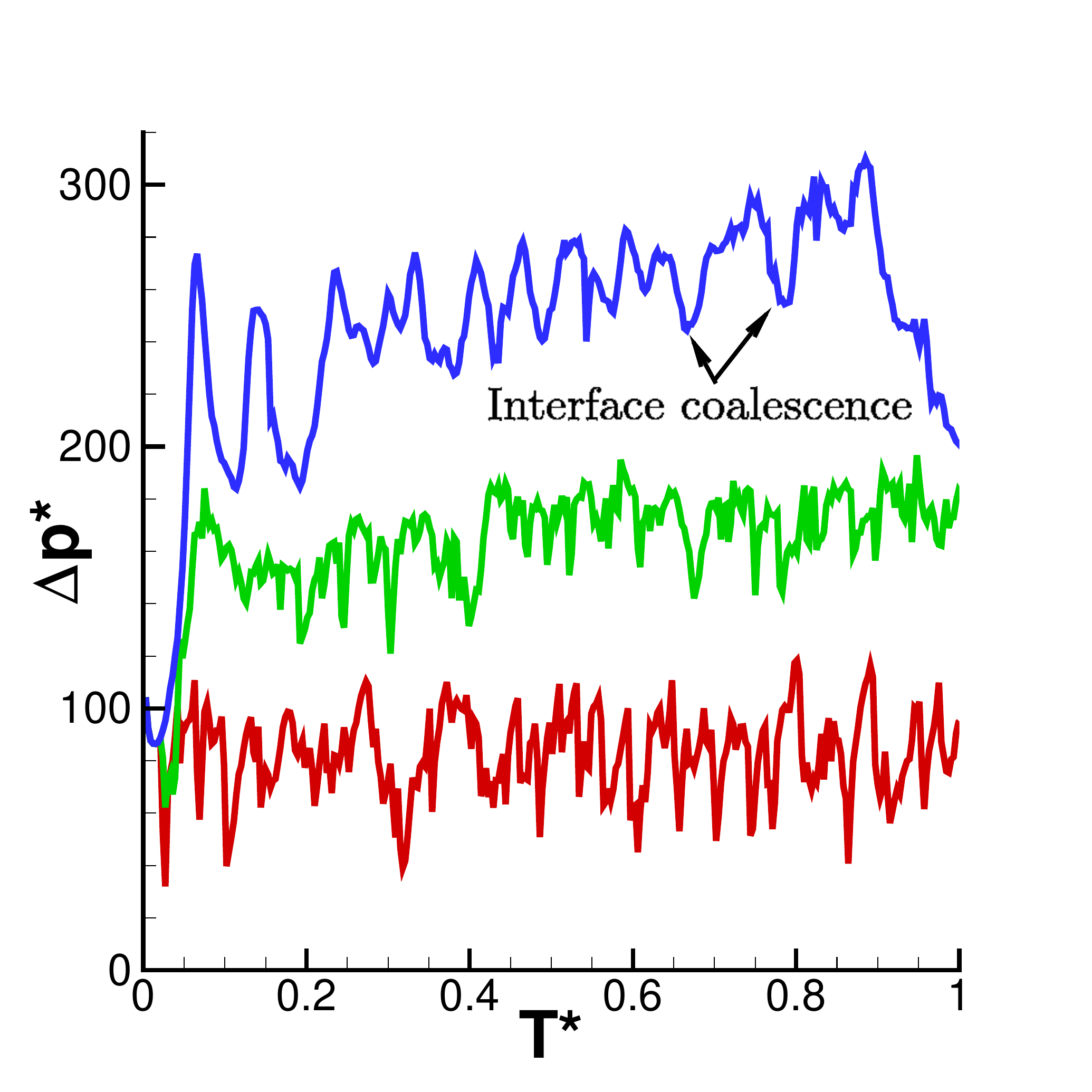}\\
		(a)&(b)&(c)
	\end{tabular}
	\caption{Driving pressure ($\Delta p^*=\frac{\Delta p\overline{D}}{\mu\abs{\vb{U}}}$) across the porous media domain, for (a) $\Ca=10^{-1}$, (b) $\Ca=10^{-2}$ and (c) $\Ca=10^{-3}$, for $\theta=\ang{30}$ (\protect\redline), $\theta=\ang{90}$ (\protect\greenline) and $\theta=\ang{150}$ (\protect\blueline). Similar to \cite{Zacharoudiou2018}, interface mergers/pore-filling events can be identified by rapid changes in the pressure signal.}
	\label{fig:pressure_drop}
\end{figure}

The transition from stable to unstable displacement is clearly visible. At small contact angles, the front advances through nearly all available channels in part due to capillary imbibition  action. As the porous media become increasingly non-wetting, i.e.~
high invading phase contact angles, the adverse capillary pressure gradient causes fingers to form, as well as pockets of defending fluid which are never displaced. The presence of Haines jumps can be detected from the pressure differential across the porous media, since occurrences are accompanied by sudden rapid decreases in local capillary pressure. The pressure signals for various $\Ca$ and $\theta$ are shown in Fig.\ \ref{fig:pressure_drop}.

\begin{figure}[]
	\begin{tabular}{ccc}
		\includegraphics[width=0.32\linewidth,trim=20mm 0 0 0]{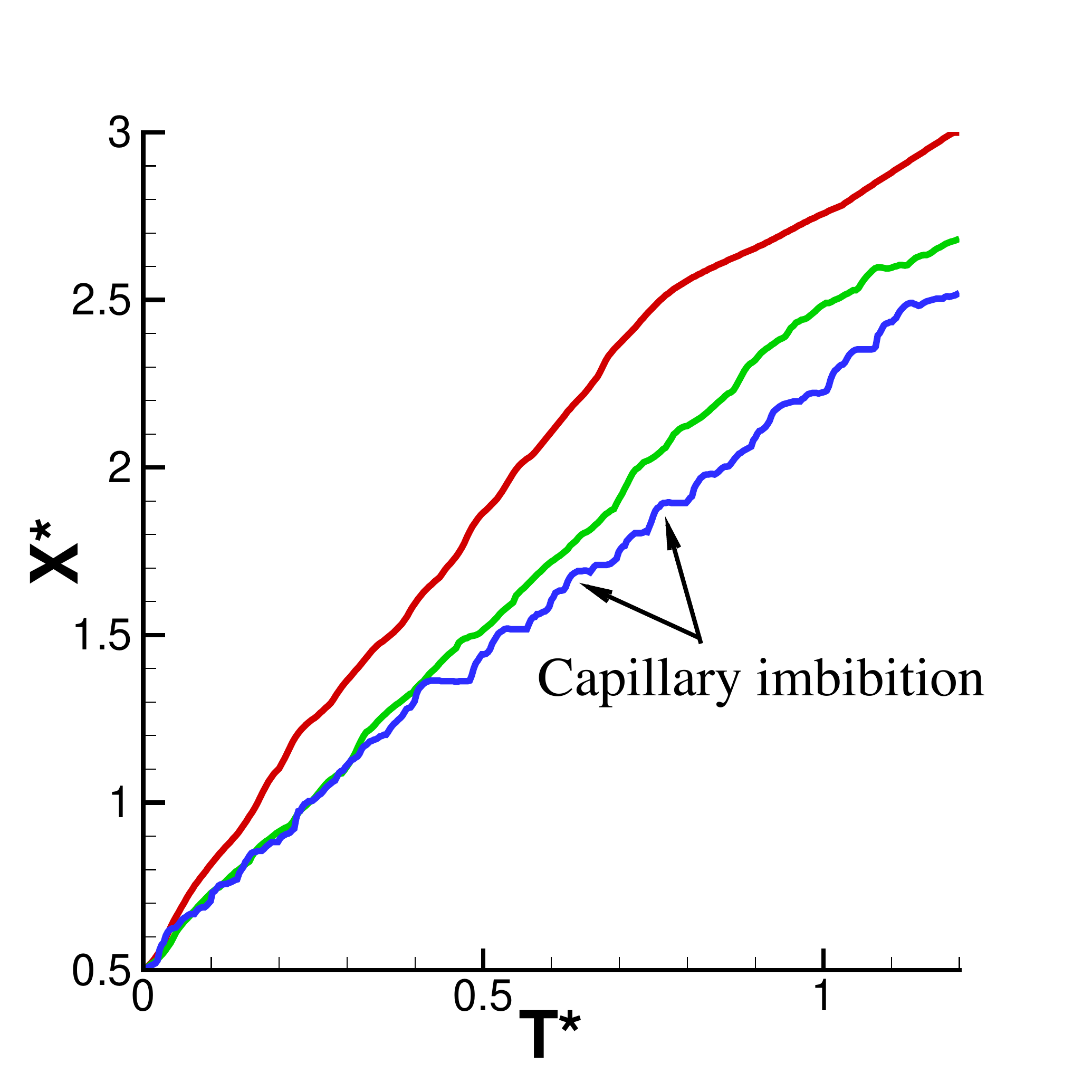}&
		\includegraphics[width=0.32\linewidth,trim=20mm 0 0 0]{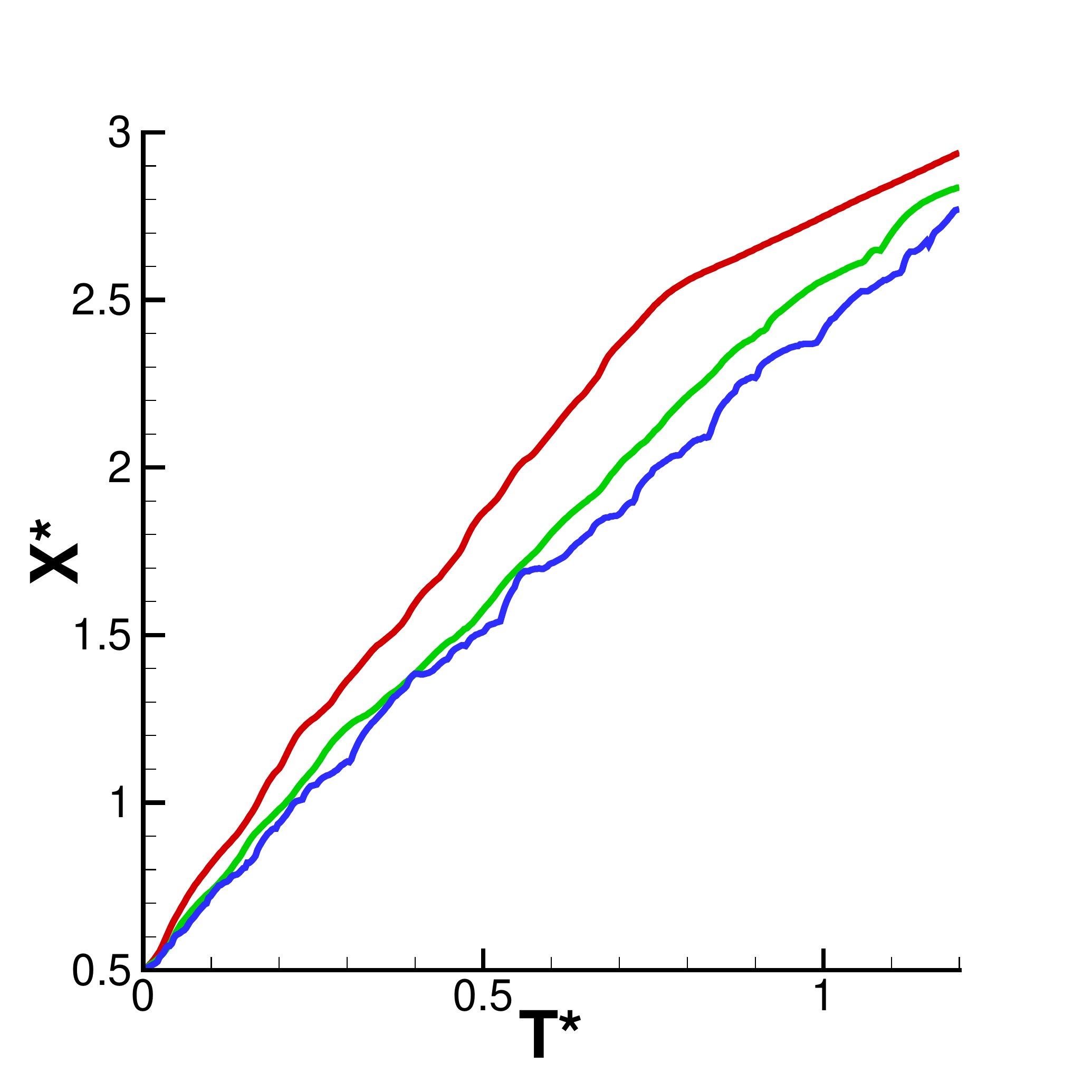}&
		\includegraphics[width=0.32\linewidth,trim=20mm 0 0 0]{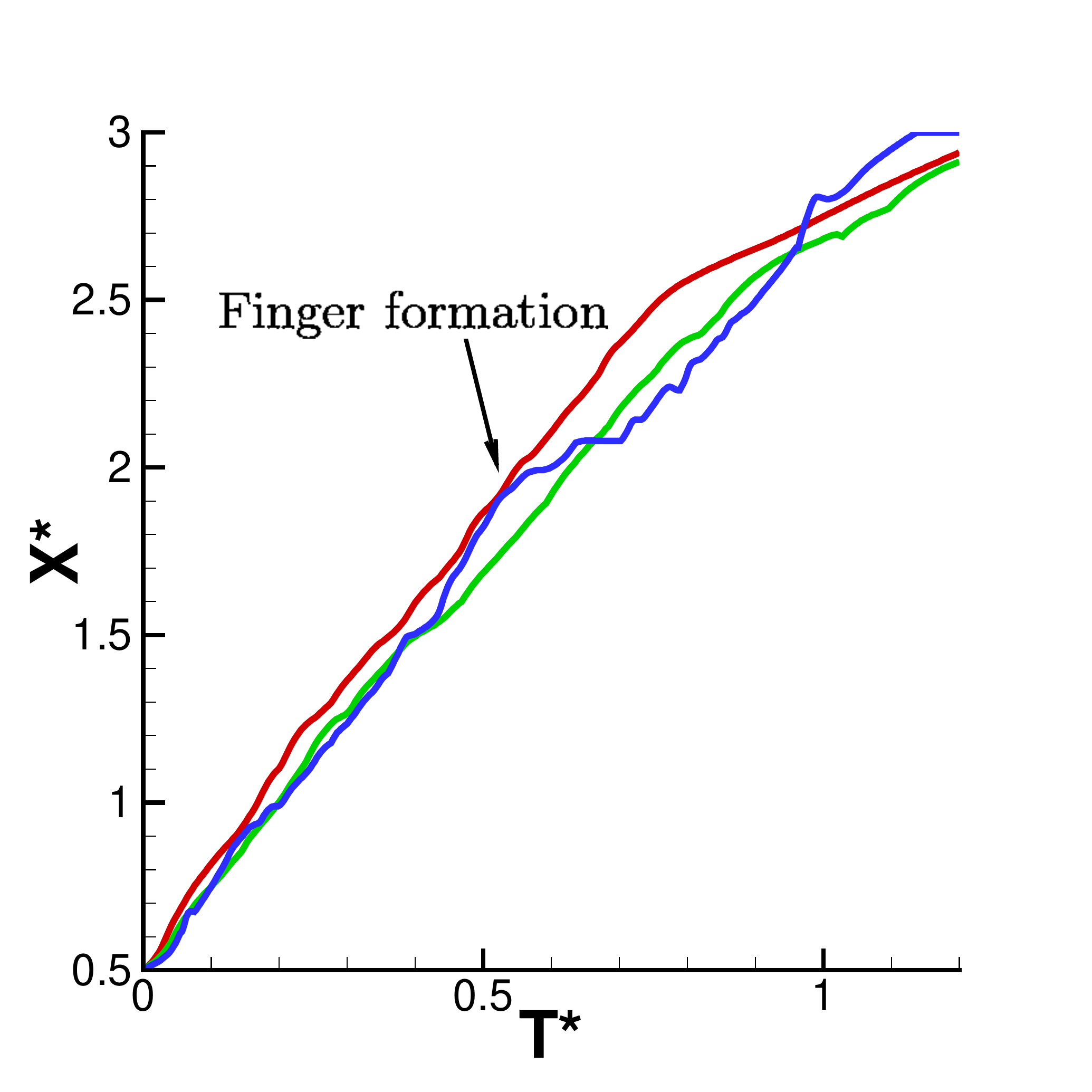}\\
		(a)&(b)&(c)
	\end{tabular}
	\caption{Comparison of front location of the invading fluid in the longitudinal direction, for (a) $\theta=\ang{30}$, (b) $\theta=\ang{90}$ and (c) $\theta=\ang{150}$, with $\Ca=10^{-1}$ (\protect\redline), $\Ca=10^{-2}$ (\protect\greenline) and $\Ca=10^{-3}$ (\protect\blueline). As the capillary number decreases, the movement of the front becomes increasingly dominated by the capillary imbibition between pores (a) and Haines jumps (c).}
	\label{fig:front_loc}
\end{figure}

\begin{figure}[]	
	\begin{tabular}{ccc}
		\includegraphics[width=0.32\linewidth,trim=20mm 0 0 0]{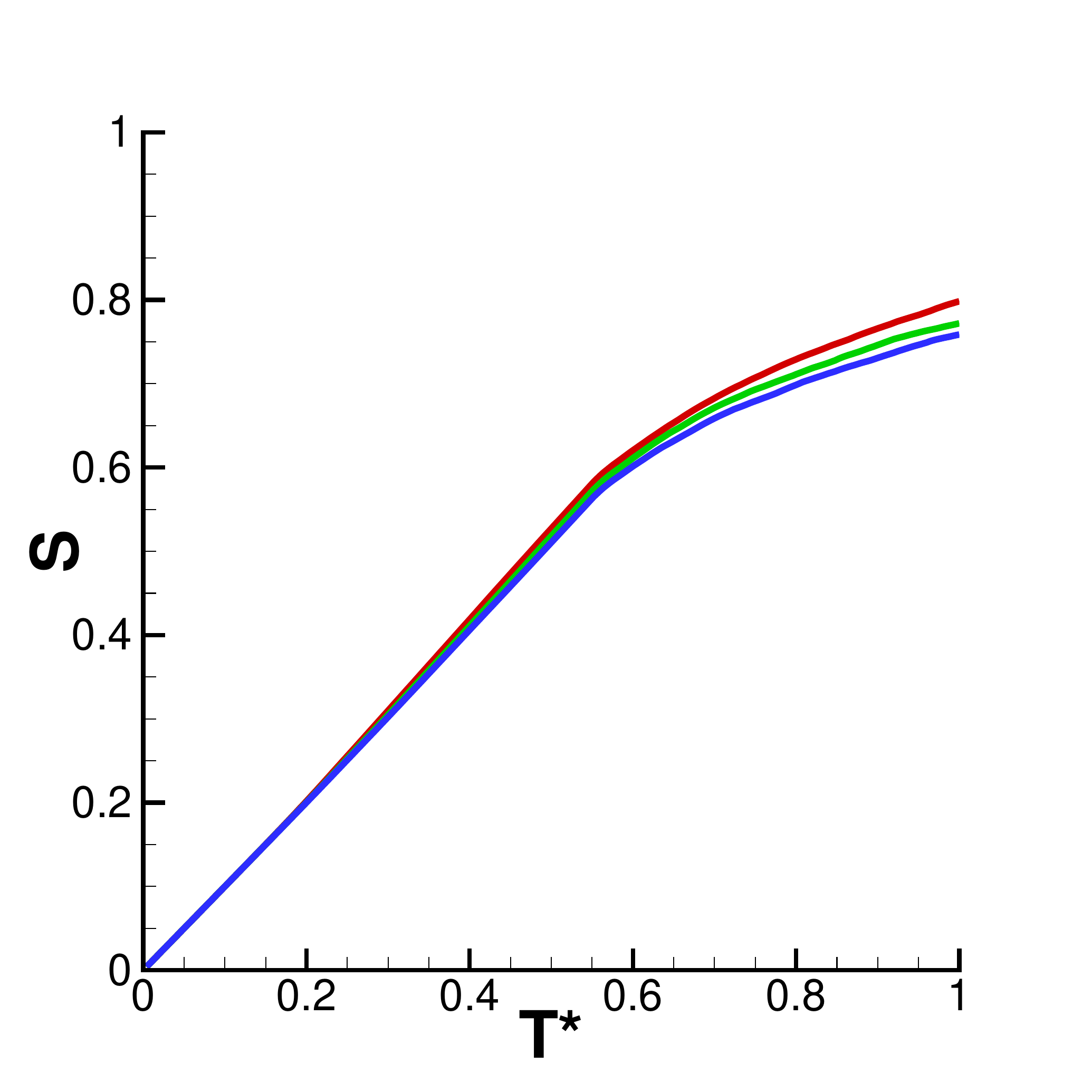}&
		\includegraphics[width=0.32\linewidth,trim=20mm 0 0 0]{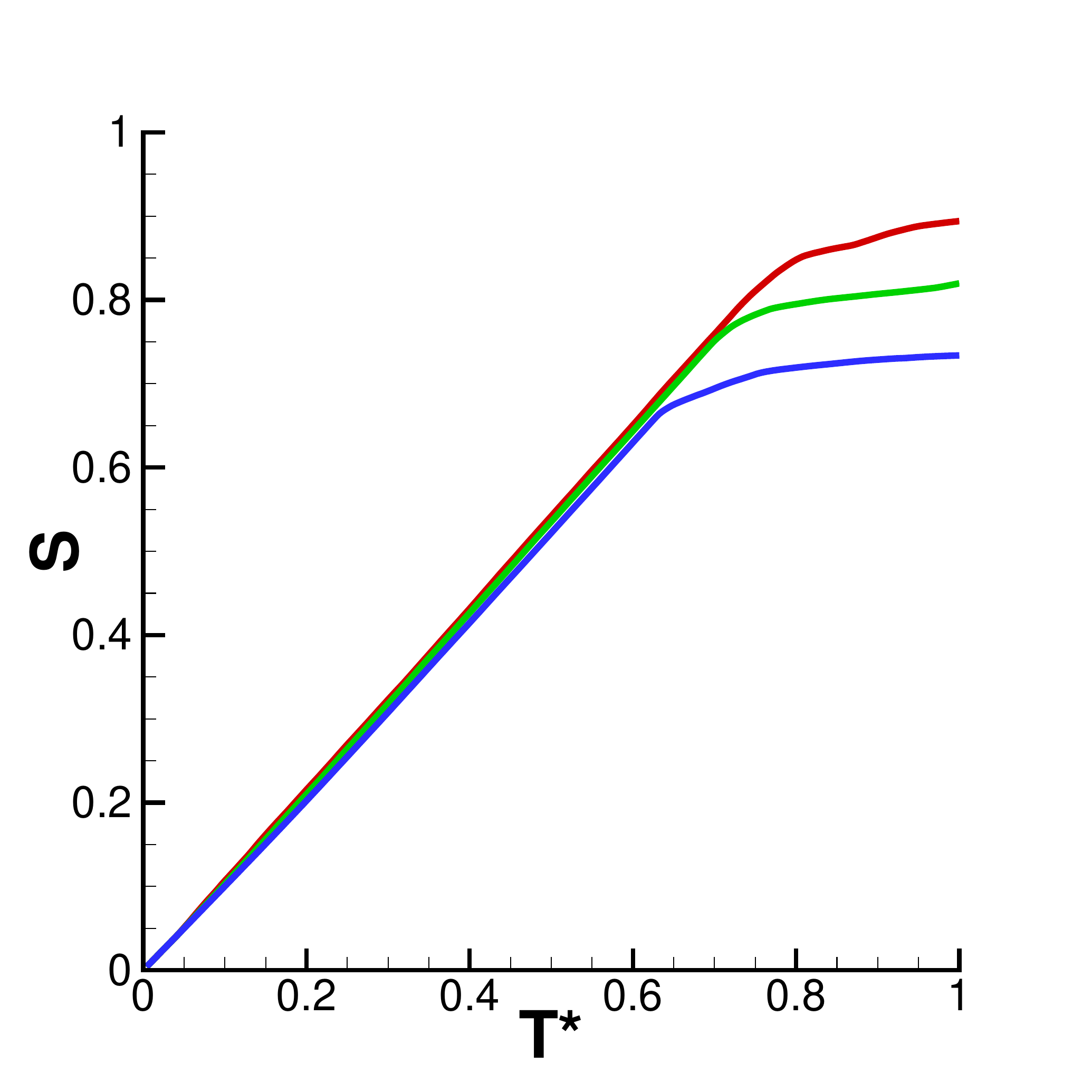}&
		\includegraphics[width=0.32\linewidth,trim=20mm 0 0 0]{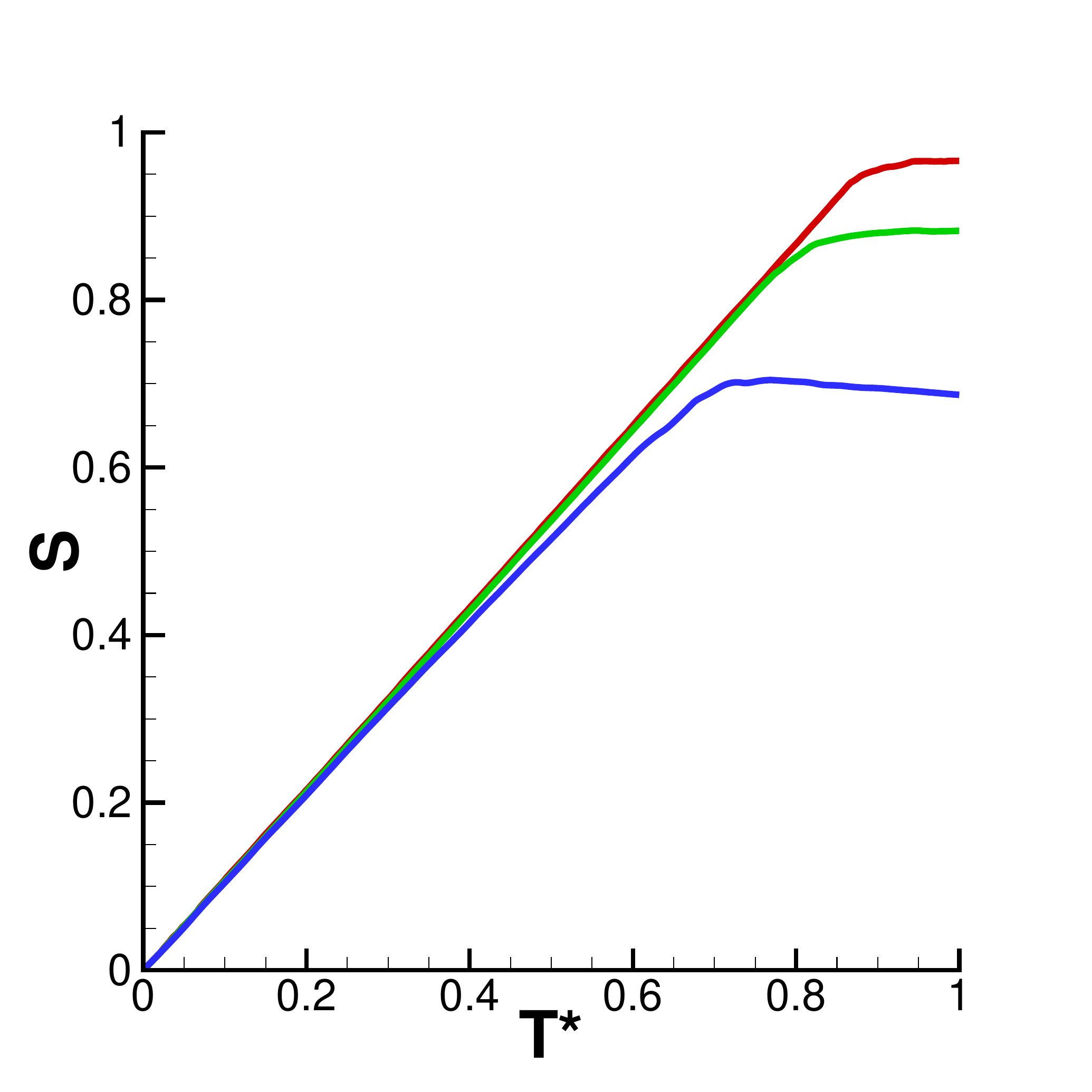}\\
		(a)&(b)&(c)
	\end{tabular}
	\caption{Displacement fraction $S$ of the invading fluid with respect to the pore volume, for (a) $\Ca=10^{-1}$, (b) $\Ca=10^{-2}$ and (c) $\Ca=10^{-3}$, with $\theta=\ang{30}$ (\protect\redline), $\theta=\ang{90}$ (\protect\greenline) and $\theta=\ang{150}$ (\protect\blueline). In (a), saturation takes longer due to the presence of capillary fingering and the formation of droplets. The kinks in (b) and (c) are due to residual fluid pockets which are not displaced.}
	\label{fig:disp_vol}
\end{figure}

At high $\Ca$, the dynamics of the flow are dominated primarily by viscous forces. As a result, the average pressure gradient remains relatively constant, and varies only slightly as the flow establishes a steady state. In the viscous regime, the dynamics of the interface are nearly independent of contact angle. As $\Ca$ is lowered, two main things are observed to occur. First, the pressure differential needed to drive the flow increases with $\theta$. For a non-wetting porous media, a high mean curvature interface forms as it travels through a throat, resulting in a large capillary force acting in the direction opposite to the flow. For a more wetting porous media, i.e.~low invading phase contact angles, the direction of the curvature is reversed, which acts to expedite the progression of the interface through the throat. The second observation is that due to the presence of Haines jumps and capillary imbibition, the intensity of pressure fluctuations increases; as we discuss next, when capillary fingering is present, the  fraction  of  the defending fluid that has been displaced from the domain decreases in this case.

The degree to which capillary forces play a role in the flow dynamics can also be depicted by examining the front propagation. In Fig.\ \ref{fig:front_loc}, the front displacement, defined as the farthest longitudinal location of the interface at each time, for different $\Ca$ at various $\theta$ is shown. For any contact angle, the front propagation is relatively unaffected when Ca is large. Decreasing the Ca results in a greater variance over time in the front propagation, due to imbibition at low Ca, and capillary fingering at higher Ca.

The dominance of capillary forces at low $\Ca$ manifests itself primarily through the unsteady propagation of the front. This is due to high-frequency front oscillations resulting from both capillary rise and Haines jumps, and lower frequency oscillations that are the result of the formation of fingers. Viscous regimes are characterized by a relatively smooth propagation of the front when compared to capillary-driven regimes.

Examining the displacement fraction, $S$, defined as the ratio between the invading fluid volume and the total fluid field volume,
can yield useful information regarding the different displacement regimes. Figs.\ \ref{fig:disp_vol} shows 
the displacement fraction $S$ as a function time for various regimes regime; as shown, a gradual increase in $S$ occurs up to $S_\text{max}$, where $S_\text{max}$ denotes the displacement fraction after steady state is obtained. In the viscous regime, the shear forces cause the interface to form long fingers which primarily follow the local velocity field. As a result, the interface does not immediately make contact with the surfaces of obstacles and form contact lines, which act to fill the pores normal to the flow direction through capillary action. As a result, the time scale on which the throats and pores normal to the flow direction are filled is larger when a contact line does not form, as the interface will only fill pores at a rate governed by the flux through pores themselves.

As $\Ca$ decreases, increasing the dominance of surface tension, and therefore capillary pressure, the interface is forced to expand in the transverse direction and fill more pores regardless of the pore orientation with respect to the flow. This effect can be seen in Figs.\ \ref{fig:disp_vol}b and \ref{fig:disp_vol}c, where at lower $\Ca$, a more pronounced initial linear displacement phase is observed, and in contrast to Fig.\ \ref{fig:disp_vol}a, the linear displacement phase can be seen to end abruptly. This is due to the sudden formation of pockets of the defending fluid resulting from Haines jumps. Only in the case of $\theta=\ang{30}$ is full displacement observed, since this regime represents a fully stable front which displaces all of the defending fluid. We also note that here we 
present an integrated quantity $S$, which is meant to illustrate the amount of fluid displaced from the porous media by the invading phase, as opposed to identifying an individual Haines jump. The displacement is most efficient for a stable front propagation compared to the extremes where viscous or capillary fingering is present. 

\section{Conclusions}
\label{sec:conclusions}

This work presents an extensive computational study of 
the dynamics of a two-phase flow in a porous media model.
Direct numerical simulations provide an enhanced understanding of the subtle and complex competition between capillarity and wetting effects 
in a porous media consisting of two immiscible fluids displacing in a two-dimensional micro-geometry filled with randomly sized and randomly distributed cylinders. We show the transition from a stable front to
a patterned displacement, with a focus on the wetting features of the porous media. The study provides detailed insight into various scenarios as we change $\Ca$ and $\theta$, while keeping the viscosity ratio fixed. We leave the study of the viscosity ratio effect to future work. We also study the Haines jump, namely rapid pore-scale displacement. Our simulations reveal the characteristic time and length scales of the Haines jump. Our numerical analysis of the Haines jump in a simple pore configuration, where cylinders of equal size are placed at the vertices of equilateral triangles, provide further insight into the effect of the contact angle at which the Haines jump is predicted. Finally, we do not address issues related to
the moving contact line problem \cite{Afkhami2018}, which is out of the scope of this study.
An extension of this work should consider the effects of slip, and validate the results with experimental measurements.
One of the main advantages of the direct numerical VOF/IBM method presented in this work is its ability to model complex geometries. Future work will consider the representation of three-dimensional domains to 
demonstrate the computational capabilities of our numerical methods with respect to other existing methods.

\end{document}